\documentclass[%
 reprint,
superscriptaddress,
 amsmath,amssymb,
 aps,
]{revtex4-2}

\usepackage{graphicx}
\usepackage{dcolumn}
\usepackage{bm}

\begin{document}
\preprint{APS/123-QED}


\title{Nodal lines in momentum space: topological invariants and recent realizations in photonic and other systems}


\author{Haedong Park}
 \affiliation{
 School of Physics and Astronomy, Cardiff University, Cardiff CF24 3AA, United Kingdom
 }
\author{Wenlong Gao} 
 \affiliation{
 Department of Physics, Paderborn University, Warburger Straße 100, 33098 Paderborn, Germany
 }
\author{Xiao Zhang} 
 \affiliation{
 School of Physics, Sun Yat-sen University, Guangzhou 510275, China
 }
\author{Sang Soon Oh}%
 \email{ohs2@cardiff.ac.uk}
\affiliation{
 School of Physics and Astronomy, Cardiff University, Cardiff CF24 3AA, United Kingdom
}%
\date{\today}
	
\begin{abstract}
Topological insulators constitute one of the most intriguing phenomena in modern condensed matter theory. 
The unique and exotic properties of topological states of matter allow for unidirectional gapless electron transport and extremely accurate measurements of the Hall conductivity. 
Recently, new topological effects occurring at Dirac/Weyl points have been better understood and demonstrated using artificial materials such as photonic and phononic crystals, metamaterials and electrical circuits. 
In comparison, the topological properties of nodal lines, which are one-dimensional degeneracies in momentum space, remain less explored. 
Here, we explain the theoretical concept of topological nodal lines and review recent and ongoing progress using artificial materials. 
The review includes recent demonstrations of non-Abelian topological charges of nodal lines in momentum space and examples of nodal lines realized in photonic and other systems. 
Finally, we will address the challenges involved in both experimental demonstration and theoretical understanding of topological nodal lines.
\end{abstract}


\maketitle
	
\section{Introduction} 
Degeneracies in energy-momentum relations, so-called band structures, play an important role in many fields of physics from classical mechanics to condensed matter physics and optics. 
Recently, the study on degenerate points in band structures opened a new path to observe many exotic topological behaviors. 
For instance, Dirac points, point degeneracies with a linear dispersion relation in two-dimensional momentum space \cite{Rechtsman_Nature_2013,Yang2015}, have been used to show chiral one-way edge/surface states in a band gap \cite{Li2018,Gong_ACSPhotonics_2020,Wong_PRR_2020,Kim_2020_LSA}. 
Weyl points, which are also point degeneracies with a linear dispersion but in three-dimensional momentum space \cite{Lu2013,Lu_2015_Science,Yang_Science_2018}, have been used to generate Fermi arc-like surface states \cite{Yang_OptExp_2017,Noh_NatPhys_2017,Guo_PRL_2019,Yang2019,He_NatComm_2020_4}.
In particular, Weyl points are very interesting and important in topological classification because they are stable meaning that the Weyl points are robust to perturbations when only one of the inversion ($\mathcal{P}$) symmetry and time-reversal ($\mathcal{T}$) symmetry is broken in three-dimensional space \cite{Lu2013}.

By recovering the broken symmetry in a structure with Dirac/Weyl points, i.e., making it $\mathcal{P}$ and $\mathcal{T}$ symmetric,  a nodal line 
\cite{Kennedy_PRB_2016,Ahn_2018_PRL,Wu2019,Park_2021a_ACSPhotonics}, a one-dimensional degeneracy, can be created.   
Nodal lines have drawn attention because they can feature two-dimensional  surface states bounded by the projected nodal lines, called drumhead surface states \cite{Rui_Yu_PRL_2015,Kim_PRL_2015,Wang_SciRep_2021}, and exhibit non-Abelian band topology \cite{Wu2019}.
Moreover, as nodal lines have higher dimensions than Dirac/Weyl points, they show various shapes, for instance, a simple nodal line, a nodal ring \cite{Gao2018,Lingbo_PRL_2019,Yang_2021_NatComm}, nodal knots \cite{Kedia_PRL_2013,Bi_2017_PRB,Lee2020} for a single nodal line or a Hopf link \cite{Xie_2019_PRB,Lee2020,Tiwari_2020_PRB,Unal_PRL_2020,Yang2020,Park_2021a_ACSPhotonics}, a nodal chain \cite{Bzdusek2016,Yan_2018_NatPhys,YangZhesen_2020_PRL_ZonesPolynomial,Park_2021a_ACSPhotonics} for multiple nodal lines. 

To study the topological characteristics of the nodal lines, the starting point is to define topological invariants which are the conserved quantities when any topological phase transitions do not occur, i.e., the topological phase do not change under certain perturbations \cite{Lu_NatPhoton_Review_2014}.      
Topological invariants of Dirac/Weyl points and nodal lines are often called \emph{topological charges} similarly to electric/magnetic charges of electric/magnetic monopoles. 
While the topological charge of a simple nodal line, that is the degeneracy between two bands, can be described by a Berry phase \cite{Berry_1984}, the definition of topological charge for multiple nodal lines becomes more complex because three or more bands are involved. 
Recently, it was shown that the topological charges of the nodal lines in a three-band system can be described by quaternion numbers, forming a non-Abelian group \cite{Wu2019} and the experimental observation of the quaternion charges has been reported \cite{Guo2021}.

Along with the progress in understanding band topology, recent years saw the rapid development of topological photonics which studies and demonstrates the topological states of photons in metamaterials (including metallic photonic crystals) \cite{Khanikaev_NatMat_2013,Gao_PRL_2015,Gao_NatComm_2016,Guo_PRL_2017,Gao2018,Yan_2018_NatPhys,Yang_2021_NatComm} and dielectric photonic crystals \cite{Lu2013,Slobozhanyuk2017,Park_2020_ACSPhotonics,Jo_2021_AppliedMaterialsToday,Park_2021a_ACSPhotonics}. 
In contrast to electronic systems these photonic materials have great advantages because one can design a structure and manipulate the propagation properties of photons with more freedom and in a wide range of frequency spectra. 
For this reason, many exciting advances have been made, for example, Weyl fermions have been demonstrated using a dielectric photonic crystal called double gyroid \cite{Lu2013,Park_2020_ACSPhotonics,Jo_2021_AppliedMaterialsToday} and unidirectional edge modes have been demonstrated using a magnetic photonic crystal \cite{Wang2009c,Yang_OptExp_2017}.
On the other hand, however, the progress on nodal lines in photonics has been rather slow compared to the study on Dirac/Weyl photonic crystals. 
The possible reasons can be summarized as follows. First, investigating nodal lines in three-dimensional momentum space requires more amount of computations than Dirac or Weyl points. Second, it was very recent that nodal chains \cite{Yan_2018_NatPhys} or nodal links \cite{Yang2020,Wang_LSA_2021,Park_2021a_ACSPhotonics} started to be realized although nodal ring by a dielectric photonic crystal was already reported in Ref.~\cite{Lu2013}. Lastly, discussions on new topological invariants of nodal lines such as non-Abelian topological charges have been made very recently \cite{Wu2019}. 
Therefore, reviewing the recent work on nodal lines is very timely and summarising important concepts is essential for more exciting outcomes that will be generated in the field of topological photonics in the near future.

In this review, we will aim to cover the basic theory of topological physics with a focus on topological nodal lines and introduce important examples of topological nodal lines demonstrated in various artificial material systems. 
In Section~\ref{section:theory}, we describe the degeneracies in band structures including Weyl points and nodal lines. 
In Section~\ref{section:Topological_invariants}, we explain Abelian and non-Abelian topological invariants using two examples, the Berry phase and the Wilczek-Zee connection. In Section~\ref{section:PhotonicExample},  we highlight examples of topological nodal lines in metamaterials, photonic crystals and photonic systems with synthetic dimensions. In Section~\ref{section:example}, we discuss examples in electronic crystals, phononic crystals and electrical circuits. 

\begin{figure}
    \includegraphics{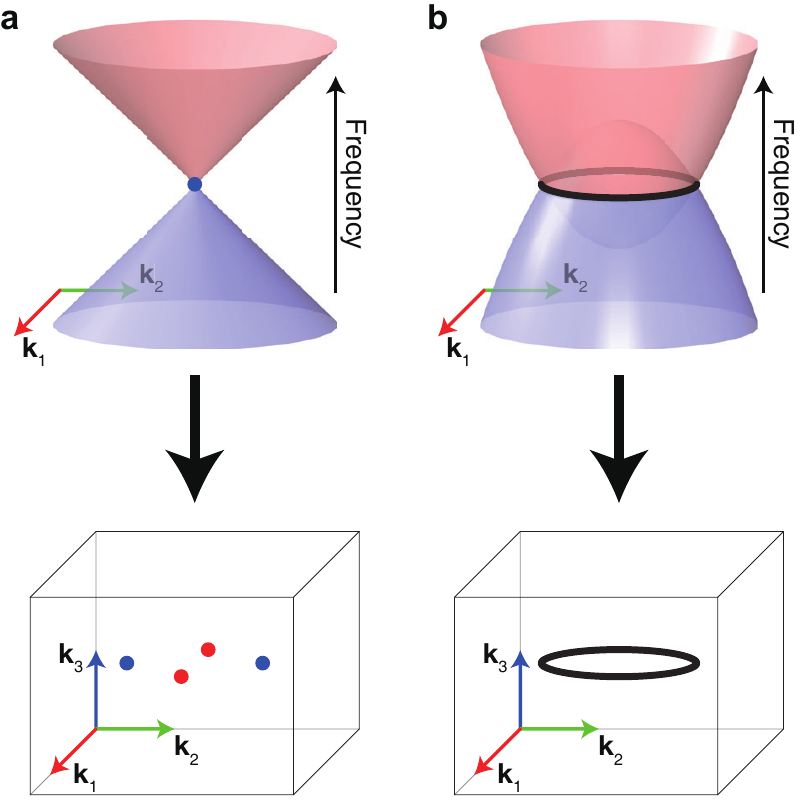}
    \caption{
        \label{fig:WeylPoint_and_NodalLine}
            (a),(b) Schematics of a Weyl point and a nodal line, respectively.
    }
\end{figure}

\section{Band degeneracies} 
\label{section:theory}
In an electronic band structure, two adjacent bands may touch each other in one or more $\mathbf k$-point(s) meaning that the two bands have the same energy but different eigenstates. 
This is `degeneracy'. 
The concept of degeneracy can also be applied to photonic/phononic band structures which are the frequency-wavevector relations for waves in a periodic array of photonic/phononic atoms. 
In three-dimensional momentum space, the degeneracies can be classified as zero-, one-, and two-dimensional depending on their dimensionality. 
In this section, we will explain zero- and one-dimensional degeneracies which can be Weyl points and nodal lines, respectively, and introduce diverse categories of nodal lines that are classified by their shapes. 

\subsection{Zero-dimensional degeneracies: Weyl points}
\label{section:WeylPoints}
A representative example of a zero-dimensional degeneracy is a Weyl point \cite{Lu2013,Lu_2015_Science,Yang_Science_2018}. 
The Weyl point was named thus because the dispersion around the degenerate point is governed by the Weyl Hamiltonian $H(\mathbf k) = v_1 k_1 \sigma_1 + v_2 k_2 \sigma_2 + v_3 k_3 \sigma_3$ where $\sigma_i$ are the Pauli matrices.
In three-dimensional momentum space, the Weyl point acts as a monopole that emits or soaks the Berry flux similar to a magnetic monopole where the magnetic flux departs or terminates. 
The mathematical definition of the Berry flux will be introduced in Section~\ref{section:BerryPhase}. 
In the Weyl Hamiltonian, the $\sigma_2$ term can exist only when one of the inversion ($\mathcal{P}$) and time-reversal ($\mathcal{T}$) symmetries is broken \cite{Lu2013}. This is a necessary condition for the existence of Weyl points. 
If we set $v_1 = v_2 = v_3 = 1$, the eigenenergies of the Weyl Hamiltonian are expressed as $E = \pm \left| \mathbf k \right|$. 
Thus, the band structure shows a point degeneracy at $\mathbf k = \mathbf 0$ 
and a linear dispersion around the degeneracy, as shown in Figure~\ref{fig:WeylPoint_and_NodalLine}a. 
The eigenstates of the Weyl Hamiltonian at $\mathbf k \neq \mathbf 0$ can be expressed as $\psi_1 = \left[ \cos \left( {\theta / 2} \right), e^{i \phi} \sin \left( {\theta / 2} \right) \right]$ and $\psi_2 = \left[ e^{-i \phi} \sin \left( {\theta / 2} \right) , -\cos \left( {\theta / 2} \right) \right]$ using the spherical coordinate system $\left( r, \theta, \phi \right)$. 
Then, the Berry curvature is expressed as $\pm 1/\left(2 k^2 \right) \hat{\mathbf r}$ where $\hat{\mathbf r}$ is the unit vector in the radial direction in momentum space \cite{Griffiths_QM_2005,Lu_2015_Science} implying that the Weyl point becomes a sink or source of the Berry flux.

In the early 2010s, significant efforts have been made to realize Weyl points \cite{Wan2011,Burkov_2011_PRL,Xu_PRL_2011,Yang2011a,Halasz_PRB_2012,Hosur_PRL_046602,Aji_PRB_2011}, 
and a double gyroid structure was theoretically proposed as a Weyl photonic crystal in 2013 \cite{Lu2013}. 
Two years after this theoretical work, Weyl points were experimentally observed in the microwave frequency range \cite{Lu_2015_Science}. 
Following this first experimental demonstration, the realization of Weyl points has been achieved using photonic crystals \cite{Yang_OptExp_2017,Luyang_PRA_2016,Goi_LaserPhotonRev_2018,Lu_NatComm_2018,Yang_Science_2018,Jia_Science_2019,Fruchart_PNAS_2018,Park_2020_ACSPhotonics,Yang_PRL_2020,Jo_2021_AppliedMaterialsToday}, 
phononic crystals \cite{Xiao_NatPhys_2015,Li2018,He_NatComm_2020_4,He_Nature_2018,Peri_NatPhys_2019,Takahashi_PRB_2019,He_NatComm_2020_5,Wang_NatComm_2021}, metals \cite{Burkov_AnnRevCondMattPhys_2018,Galitski_PRL_2018,Mizobata_PRB_2020,Sorn_PRB_2021} and semimetals \cite{Soluyanov2015,Sie_Nature_2019,Heidari_PRB_2020,Ilan_NatRevPhys_2020,Kim_PRL_2017,Wang_PRL_2020,Yuan_NatComm_2018}.

\begin{figure*}[!ht]
    \centering
    \includegraphics{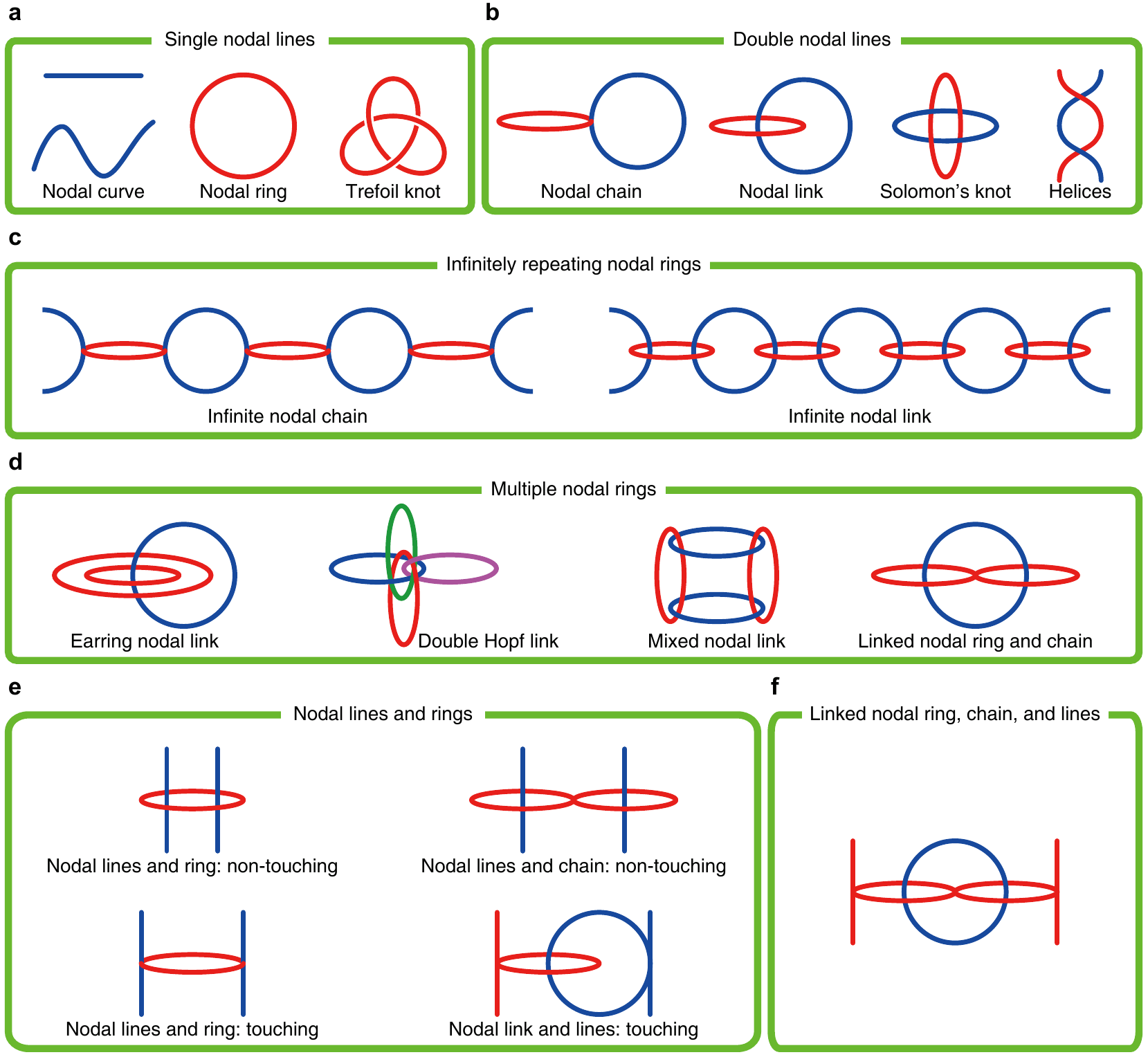}
    \caption{
        \label{fig:Classification_of_NodalLine}
            Classification of nodal lines.
    }
\end{figure*}
\subsection{One-dimensional degeneracies: nodal lines}
\label{section:NodalLines}
One-dimensional degeneracies (co-dimension $N-1$ in a $N$-dimensional space), called `nodal lines’, can be found when two bands touch each other on a line in momentum space  \cite{Burkov2011,Kennedy_PRB_2016,Ahn_2018_PRL,Wu2019,Park_2021a_ACSPhotonics}, as illustrated in Figure~\ref{fig:WeylPoint_and_NodalLine}b. 
The nodal lines can be classified by their shapes and the connectivities between other nodal lines, as shown in Figure~\ref{fig:Classification_of_NodalLine}.

As shown in  Figure~\ref{fig:Classification_of_NodalLine}a, a nodal line can have a loop shape to form a nodal ring \cite{Lu2013,Fang_PRB_2015,Weng_PRB_2015,Chen_NatComm_2015,Xie_APLMat_2015,Ezawa_PRL_2016,Zhao_PRB_2016,Chan_PRB_2016,Li_PRL_2016,Bian_NatComm_2016,Gao2018,Nomura_PRMat_2018,Deng2019,Lingbo_PRL_2019,Tiwari_2020_PRB,Yang_2021_NatComm,Li_2021_PRB}
or knot, such as a trefoil knot \cite{Kedia_PRL_2013,Bi_2017_PRB,Lee2020}, a double trefoil knot \cite{Bi_2017_PRB,Ezawa_PRB_2017}, a cinquefoil knot \cite{Bi_2017_PRB,Kedia_PRL_2013}, or a figure 8-knot \cite{Lee2020}.
Such knots cannot be transformed into a nodal ring without cutting or intersecting them.

If two rings intersect each other, they form a nodal chain \cite{Yan_PRB_2017,Sun_PRL_2018,Gong_PRL_2018,Zhou_PRB_2018,Belopolski2019,Merkel_CommPhys_2019,Chang_2017_PRL}.
If two rings are tied without touching so that they cannot be separated by cutting or passing them, they form a nodal link \cite{Wilczek_PRL_1983,Moore_PRL_2008,Neupert_PRB_2012,Kedia_PRL_2013,Deng_PRB_2013,Deng_PRB_2014,Yan_PRB_2017,Chang_2017_PRL,Ezawa_PRB_2017,Liu_PRB_2017,Lian_PRB_2017,Zhou_PRB_2018,Xie_2019_PRB,Lee2020,Tiwari_2020_PRB,Unal_PRL_2020} or Solomon's knots \cite{Ezawa_PRB_2017,Bi_2017_PRB} (Figure~\ref{fig:Classification_of_NodalLine}b). Each ring of the nodal chain/link can be formed by the same or a different pair of bands. Especially, if the two rings of the nodal chain originate from different set of bands in a three-band system, the three bands meet at a single point where the nodal rings touch \cite{Wu2019,Lenggenhager_2021_PRB,SPark_2021_arXiv,Lange_2021_arXiv}. This is called a triple point.

Multiple nodal rings may form an infinite nodal chain \cite{Bzdusek2016,Wang_NatComm_2017,Yan_2018_NatPhys,Gong_PRL_2018,YangZhesen_2020_PRL_ZonesPolynomial,Park_2021a_ACSPhotonics} or a link \cite{Xie_2019_PRB,Lenggenhager_2021_PRB,Park_2021a_ACSPhotonics} due to the periodicity of momentum space (Figure~\ref{fig:Classification_of_NodalLine}c). Although Figure~\ref{fig:Classification_of_NodalLine}c shows only a one-dimensional infinite chain and a link, they can also form a two- or three-dimensional infinite chain \cite{Yan_2018_NatPhys} or a link \cite{Xie_2019_PRB}.

In some cases, multiple nodes of different types can appear mixed.
First, the mixed shape of the nodal rings appear as earring nodal links \cite{Wu2019,Tiwari_2020_PRB}, multiple Hopf links \cite{Kedia_PRL_2013,Ezawa_PRB_2017,Zhou_PRB_2018,He_2020_PRA}, mixed nodal links \cite{Tiwari_2020_PRB}, and the linked nodal ring and a chain (Figure~\ref{fig:Classification_of_NodalLine}d) \cite{Yang2020,Wang_2021_arXiv}. 
Second, the nodal lines and a nodal ring/chain can be linked to show the non-touching between nodal lines and rings \cite{Tiwari_2020_PRB,Yang2020,SPark_2021_arXiv}.
Inversely, the nodal lines and nodal ring/link can be chained to show the touching between nodal lines and ring \cite{Bzdusek2016,Gong_PRL_2018,Lu_PRA_2020,Wang_2021_arXiv} or the touching between nodal lines and link \cite{Wu2019,Wang_2021_arXiv}, as shown in Figure~\ref{fig:Classification_of_NodalLine}e. 
Finally, if the linked nodal ring and chain \cite{Wang_LSA_2021} and the nodal lines touch, the linked nodal ring, chain, and lines \cite{Yang2020} are generated  (Figure~\ref{fig:Classification_of_NodalLine}f).

For the above classification, we make one assumption for brevity. As we will look at a periodic system, the momentum space is periodic. When a nodal line crosses a Brillouin zone boundary, it intersects at the same point at the opposite Brillouin zone boundary. 
Therefore, a nodal line and a double helix (Figure~\ref{fig:Classification_of_NodalLine}b) \cite{YangZhesen_2020_PRL_ZonesPolynomial,Kennedy_PRB_2016,Sun_PRL_2017,Chang_PRB_2017,Tan_APL_2018,Chen_PRB_2017,Unal_PRRes_2019,Wang_Nature_2021} can be considered as a nodal ring and a nodal link, respectively. 
In the above, however, we assumed that the first Brillouin zone is distinct from the neighboring Brillouin zones, which means that a nodal line that crosses the boundary extends from negative infinity to positive infinity.

\section{Topological invariants} 
\label{section:Topological_invariants}
The physical behaviors of bands in momentum space are interpreted using several kinds of topological invariants.
The topological invariant is a quantized number that characterizes the topological status of a given system, and the Chern number or Berry phase are examples of the topological invariants.  
The topological invariants are related to the various phenomena (e.g., surface states) of the topological insulators, which act as an insulators in their bulk and permit the electronic/photonic/phononic waves on their boundaries.
For example, in $\mathcal{T}$-symmetry broken topological insulators, one-way surface states are formed at the interface of two band-gapped materials due to the difference in Chern numbers between two bands \cite{Raghu2008, Wang2009c,Kruthoff2017,Lu_NatPhoton_Review_2014}. 
In pseudo-$\mathcal{T}$-symmetry broken or $\mathcal{P}$-symmetry broken topological insulators, surface states exist due to the difference in the spin-Chern numbers or valley-Chern numbers \cite{Wu2015a,Slager2013,Saba2020}.
In addition to the topological invariants of these gapped phases, several efforts also have been carried out to understand relations between surface states and the topological invariants of gapless phases (band degeneracies), such as Dirac points \cite{Jiang2021}, Weyl points \cite{Lu2013,Chen_NatComm_2016} or nodal lines \cite{Wu2019}.
Thus, describing the topological invariants of the band degeneracy is an important step to understanding the degeneracy and finding appropriate applications.

In the following, we review Abelian and non-Abelian topological invariants. We introduce the Berry phase \cite{Berry_1984} as an example of the Abelian topological invariants. We also explain the Wilczek-Zee phase \cite{Wilczek_PRL_1984}, which is the starting point toward the non-Abelian topological quaternion charges \cite{Wu2019}. 
Simple examples of the quaternion charges are shown, and the use of correlation vectors in full-vector field systems is explained.
Finally, the patch Euler patch class is explained with an example of non-Abelian topology. 
For more information, refer to Ref.~\cite{Berry_1984} for the Berry phase, Ref.~\cite{Wilczek_PRL_1984} for the Wilczek-Zee connection, Ref.~\cite{Wu2019} for the non-Abelian quaternion charges, and Ref.~\cite{Bouhon2020} for the Euler class derived using the Wilson loop.

\subsection{Abelian and non-Abelian topological invariants}
\label{section:Abelian_and_non_Abelian}
As noted in Section~\ref{section:theory}, a Weyl point and a nodal line are zero- and one-dimensional degeneracies, respectively. 
In fact, a nodal line in three-dimensional space has the similar features  to a Dirac point \cite{Rechtsman_Nature_2013,Yang2015,Lu_NatPhys_2016,Slobozhanyuk2017,Jin_PRL_2017,Abbaszadeh_PRL_2017,Brendel_PNAS_2017,Yang_Nature_2019,Wen_NatPhys_2019,Gong_ACSPhotonics_2020,Liu_NatComm_2020,Shao_NatNanotech_2020,Wong_PRR_2020} which is a point degeneracy in two-dimensional momentum space. 
The nodal line and the Dirac point have the same co-dimension $N-1$
and commonly correspond to $H(\mathbf k) = v_1 k_1 \sigma_1 + v_3 k_3 \sigma_3$ that does not have the $\sigma_2$ term compared to the Weyl Hamiltonian mentioned in Section~\ref{section:WeylPoints}.
In addition, calculating the topological invariant of a nodal line and a Dirac point starts from considering a closed loop around the degeneracies, whereas calculating the topological charge of a Weyl point is associated with the Berry flux on the surface enclosing the Weyl point.

In case of the Abelian charge of nodal lines, to describe the topological nature of multiple degeneracies between the same pair of bands, e.g., in a two-band system, the topological invariants are obtained by simply summing up the invariants of all the degeneracies. However, such an invariant cannot express the full topological nature of the multi-band systems. For example, when the Abelian charges are used, the nodal lines between the first and second bands and between the second and third bands commonly exhibit a topological charge of $\pm \pi$. Thus, this invariant cannot distinguish which bands make the degeneracy, and the relation of the charges between a different pair of bands cannot be described.

The non-Abelian band topology gives a solution for the multi-band systems \cite{Wu2019}. Degeneracies by a different pair of bands have different topological charges, the quaternion numbers. The mutual interaction between the different pairs of bands can be written clearly, and the topological charges satisfy the anticommutative relation.

\subsection{Berry phase}
\label{section:BerryPhase}
Berry phase is a geometrical phase that is obtained by a system when it moves along a closed path in a parameter space \cite{Berry_1984}. 
The Berry phase is path-dependent and is useful in studying the topology of the parameter space by providing a way to calculate topological invariants. 
The mathematical description of the Berry phase starts with a Hamiltonian that depends on time-varying parameters $\mathbf k = \left[ k_1 , k_2 ,\cdots \right]$, i.e., $H = H \left( \mathbf k \left( t \right) \right)$. 
We denote the orthonormal bases of $H \left( \mathbf k \left( t \right) \right)$ as $\left| u_n \left( \mathbf k \left( t \right) \right) \right\rangle$:
\begin{equation}
    H \left( \mathbf k \left( t \right) \right)
    \left| u_n \left( \mathbf k \left( t \right) \right) \right\rangle
    =
    E_n \left( \mathbf k \left( t \right) \right)
    \left| u_n \left( \mathbf k \left( t \right) \right) \right\rangle
    \label{eqn:eigenvalue} ,
\end{equation}
\begin{equation}
    \left\langle u_m \left( \mathbf k \left( t \right) \right)
    \middle|
    u_n \left( \mathbf k \left( t \right) \right) \right\rangle
    =
    \delta_{mn} .
    \label{eqn:orthonormality}
\end{equation}
Here, we assume that the eigenstates in Eqs.~(\ref{eqn:eigenvalue})-(\ref{eqn:orthonormality}) are given by not only $\left| u_n \left( \mathbf k \left( t \right) \right) \right\rangle$ but also $e^{i \gamma_n \left( t \right)} \left| u_n \left( \mathbf k \left( t \right) \right) \right\rangle$.
A state $\left| \psi \left( t \right) \right\rangle$ satisfying the time-dependent Schr\"{o}dinger equation is considered. If the parameter $\mathbf k$ changes adiabatically (that is, $\mathbf k$ varies slowly with time), the state $\left| \psi \left( t \right) \right\rangle$ that was in $n$-th state at $t = 0$ remains in the same state at $t = T$ where $T$ is the period of the cycle \cite{Kato_JPhysSocJpn_1950,Messiah_QM_1962}. The state $\left| \psi \left( t \right) \right\rangle$ is given by \cite{Berry_1984}
\begin{equation}
    \left| \psi \left( t \right) \right\rangle
    =
    e^{
        - \frac i \hbar
        \int _0 ^t dt' E_n \left( \mathbf k \left( t' \right) \right)
    }
    e^{i \gamma_n \left( t \right)}
    \left| u_n \left( \mathbf k \left( t \right) \right) \right\rangle
\end{equation}
where the first exponential term is the dynamical phase factor. By substituting $\left| \psi \left( t \right) \right\rangle$ into the time-dependent Schr\"{o}dinger equation, the geometric phase $\gamma_n$ can be obtained as an integral form. If a closed loop $\Gamma$ in $\mathbf k$-space is considered such that $\mathbf k \left( 0 \right) = \mathbf k \left( T \right)$, we have
\begin{equation}
    \gamma_n
    =
    \oint_{\Gamma} {\mathbf A}_n \cdot d {\mathbf k}
    \label{eqn:BerryPhase}
\end{equation}
where
\begin{equation}
    {\mathbf A}_n \left( \mathbf k \right)
    =
    i
    \left\langle
    u_n \left( \mathbf k \right)
    \middle|
    \nabla_{\mathbf k}
    u_n \left( \mathbf k \right)
    \right\rangle .
    \label{eqn:BerryConnection}
\end{equation}
As we consider a closed loop, the phase difference $\zeta \left( \mathbf k \left( T \right) \right) - \zeta \left( \mathbf k \left( 0 \right) \right)$ for the gauge transformation $\left| u_n \left( \mathbf k \right) \right\rangle \rightarrow e^{i \zeta \left( \mathbf k \right)} \left| u_n \left( \mathbf k \right) \right\rangle$ should be an integer multiple of $2 \pi$, so that $\gamma_n$ in Eq.~(\ref{eqn:BerryPhase}) becomes gauge-invariant. Here, $\gamma_n$ and $\mathbf A_n \left( \mathbf k \right)$ are called Berry phase and Berry connection, respectively \cite{Berry_1984,Xiao_RMP_2010,Griffiths_QM_2005}. The Berry flux (mentioned in Section~\ref{section:WeylPoints}) is given by $\iint {\mathbf F}_n \cdot d^2 \mathbf k$ where ${\mathbf F}_n \left( \mathbf k \right) = \nabla_{\mathbf k} \times {\mathbf A}_n \left( \mathbf k \right)$ is called the Berry curvature \cite{Lu_NatPhoton_Review_2014}. Regarding the closed loop $\Gamma$, a more practical form of Eqs.~(\ref{eqn:BerryPhase}) and (\ref{eqn:BerryConnection}) is
\begin{equation}
    \gamma_n
    =
    \oint_{\Gamma} i
    \left\langle
    u_n \left( \mathbf k \right)
    \middle|
    \frac \partial {\partial k}
    u_n \left( \mathbf k \right)
    \right\rangle
    d k .
    \label{eqn:BerryPhase_Practical}
\end{equation}

From the orthonormality relation $\left\langle u_m \middle| u_n \right\rangle = \delta_{mn}$, we have $\left\langle \partial u_m / \partial k \middle| u_n \right\rangle + \left\langle u_m \middle| \partial u_n / \partial k \right\rangle = 2 \operatorname{Re} \left( \left\langle u_m \middle| \partial u_n / \partial k \right\rangle \right ) = 0$. Thus, the integrand in Eq.~(\ref{eqn:BerryPhase_Practical}) is purely imaginary, and $\gamma_n$ is real. If $\left| u_n \right\rangle$ consists of only real numbers, $\gamma_n$ becomes zero. In other words, to get a non-zero Berry phase, $\left| u_n \right\rangle$ should consist of complex numbers having one or more non-zero imaginary components. One representative example for this case is the Weyl Hamiltonian with $v_1 = v_2 = v_3 = 1$, mentioned in Section~\ref{section:WeylPoints}. The Weyl Hamiltonian has $\sigma_2$ so that its eigenstates consist of complex numbers. Possessing this $\sigma_2$ term corresponds to the $\mathcal{P} \mathcal{T}$-symmetry breaking in three-dimensional space and is a necessary condition for the existence of Weyl points \cite{Lu2013}.

\subsection{Wilczek-Zee connection}
Nodal lines are generated when both $\mathcal{P}$ and $\mathcal{T}$ symmetries are conserved. Such a situation corresponds to the lack of $\sigma_2$ in the Weyl Hamiltonian. In this case, the Berry phase in Eq.~(\ref{eqn:BerryPhase_Practical}) becomes zero because one can choose a gauge that keeps the Hamiltonian and its eigenvectors real. 
Therefore, it is useful to define a non-vanishing topological invariant, for example, 
the Wilczek-Zee connection \cite{Wilczek_PRL_1984} which will be explained in the following.

A state $\left| \eta_m \left( t \right) \right\rangle$ of the $m$-th band satisfying the Schr\"{o}dinger equation can be expressed as a linear combination of the basis $\left| u_n \left( \mathbf k \left ( t \right ) \right) \right\rangle$ in Eqs.~(\ref{eqn:eigenvalue})~(\ref{eqn:orthonormality}):
\begin{equation}
    \left| \eta_m \left( t \right) \right\rangle
    =
    \sum _n W_{mn} \left| u_n \right\rangle
\end{equation}
If we assume that $\left| \eta_m \left( t \right) \right\rangle$ remain normalized, we have
\begin{equation}
    0 = \left\langle \eta_l \middle| \frac {\partial \eta_m} {\partial t} \right\rangle
    = \left\langle \eta_l \right|
    \sum_n \left[
    \frac {\partial W_{mn}} {\partial t} \left| u_n \right\rangle + W_{mn} \frac {\partial \left| u_n \right\rangle} {\partial t}
    \right] .
\end{equation}
This equation can be arranged as follows:
\begin{equation}
    \frac {\partial \mathbf W} {\partial t} \left| \mathbf u \right\rangle
    =
    - \mathbf W \frac {\partial \left| \mathbf u \right\rangle} {\partial t}
\end{equation}
or
\begin{equation}
    \left[
    {\mathbf W}^{-1} \frac {\partial \mathbf W} {\partial t}
    \right]
    \left| \mathbf u \right\rangle
    =
    - \frac {\partial \left| \mathbf u \right\rangle} {\partial t}
\end{equation}
where $\left| \mathbf u \right\rangle = \left[ \left| u_1 \right\rangle, \left| u_1 \right\rangle, \cdots \right]^T$ and $\left[ \mathbf W \right]_{mn} = W_{mn}$. Using the orthonormality condition in Eq.~(\ref{eqn:orthonormality}), we have the following skew-symmetric matrix:
\begin{equation}
    \left[
    {\mathbf W}^{-1} \frac {\partial \mathbf W} {\partial t}
    \right]_{mn}
    = - \left\langle u_n \middle| \frac {\partial u_m} {\partial t} \right\rangle
    = \left\langle u_m \middle| \frac {\partial u_n} {\partial t} \right\rangle
    = A_{mn} .
    \label{eqn:skewsymmWZ_1}
\end{equation}
Then the path-ordered integrals of this can be expressed as
\begin{equation}
    \mathbf W = \operatorname{exp} \int _0 ^T \mathbf A \left( t' \right) d t' ,
\end{equation}
or if we consider a closed loop $\Gamma$, and if we recall that $\partial u_n / \partial t = \left( \nabla_{\mathbf k} u_n \right) \cdot \left( \partial {\mathbf k} / \partial t \right)$, we can get the Wilson loop
\begin{equation}
    \mathbf W
    = \operatorname{exp} \left\{ \oint_{\Gamma} \mathbf A \left( \mathbf k \right) \cdot d {\mathbf k} \right\}
    = \operatorname{exp} \left\{ \oint_{\Gamma} \mathbf A \left( k \right) d k \right\} .
    \label{eqn:WilsonLoop}
\end{equation}
Here, the component of $\mathbf A \left( \mathbf k \right)$ is given by
\begin{equation}
    A_{mn}
    = \left\langle u_m \middle| \nabla_{\mathbf k} u_n \right\rangle
    = \left\langle u_m \middle| \frac {\partial u_n} {\partial k} \right\rangle
    \label{eqn:WZConnection}
\end{equation}
and is called the Wilczek-Zee connection \cite{Wilczek_PRL_1984}.

\begin{figure*}
    \centering
    \includegraphics{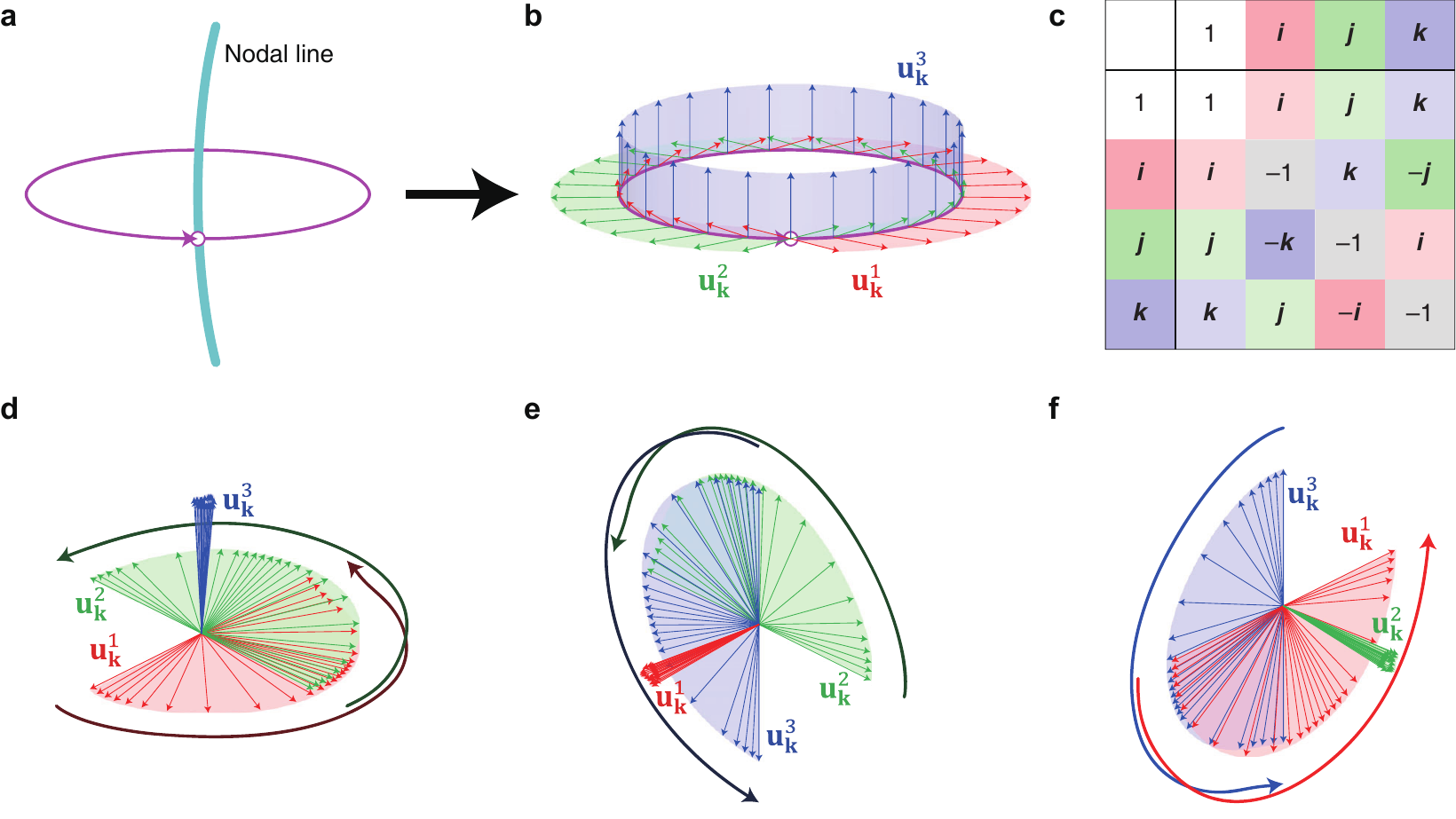}
    \caption{
        \label{fig:Schematic_Charge_IJK}
            Schematics of the non-Abelian quaternion charges. (a) A nodal line and a closed loop encircling the nodal line. (b) An example of the eigenstates along the closed loop. (c) Quaternion multiplication table. (d),(e),(f) Eigenstates gathered at the origin indicating the topological charges ${\boldsymbol k}$, ${\boldsymbol i}$, and $+{\boldsymbol j}$, respectively.
    }
\end{figure*}
\subsection{Three-band system}
In this section, the expressions derived in the previous section are applied to a three-band system.
Following formalism will be the basis of the rigorous description of quaternion charges \cite{Wu2019}.

To calculate the topological charges of an arbitrary nodal line, first, a closed loop $\Gamma \left( \alpha \right)$ ($\alpha \in \left(0, 2\pi \right]$) around the nodal line is considered (see Figure~\ref{fig:Schematic_Charge_IJK}a). 
From the eigenstates of the $m$- and $n$-th bands ($m,n=1,2,3$), the Wilczek-Zee connection (Eq.~(\ref{eqn:WZConnection})) is rewritten as follows:
\begin{equation}
    \mathbf A \left( k \right)
    =
    \left[
        \begin{array}{ccc}
            A_{11} & A_{12} & A_{13} \\
            A_{21} & A_{22} & A_{23} \\
            A_{31} & A_{32} & A_{33} \\
        \end{array}
    \right].
\end{equation}
Eq.~(\ref{eqn:skewsymmWZ_1}) leads to the skew-symmetric $\mathbf A \left( \mathbf k \right)$:
\begin{equation}
    \mathbf A \left( k \right)
    =
    \left[
        \begin{array}{ccc}
            0 & A_{12} & -A_{31} \\
            -A_{12} & 0 & A_{23} \\
            A_{31} & -A_{23} & 0 \\
        \end{array}
    \right]
    =
    {\boldsymbol \beta} \left( k \right) \cdot {\mathbf L}
    \label{eq_A_BWZ_skewsymm}
\end{equation}
where ${\boldsymbol \beta} = \left[ -A_{23}, -A_{31}, -A_{12} \right]$ and
$\left( L_i \right)_{jk} = -\epsilon_{ijk}$ \cite{Wu2019,Yang2020}.

If the closed loop encircles the nodal line formed by the bands $m=1$ and $n=2$, $\mathbf A \left( \mathbf k \right)$ is rewritten as
\begin{eqnarray}
    \mathbf A \left( k \right)
    &=&
    \left[
        \begin{array}{ccc}
            0 & A_{12} & 0 \\
            -A_{12} & 0 & 0 \\
            0 & 0 & 0 \\
        \end{array}
    \right]
    \nonumber\\
    &=&
    {\boldsymbol \beta}_{12} \left( k \right) \cdot {\mathbf L} = -A_{12} L_3 ,
    \label{eq_A_BWZ_Mat_12}
\end{eqnarray}
where ${\boldsymbol \beta}_{12} \left( k \right) = \left[ 0,0, -A_{12} \right]$.
Substituting Eq.~(\ref{eq_A_BWZ_Mat_12}) into Eq.~(\ref{eqn:WilsonLoop}) gives
\begin{eqnarray}
    W
    &=&
    \operatorname{exp}
    \left\{
        \oint_{ \Gamma(\alpha) } {\boldsymbol \beta}_{12} \left( k \right) \cdot {\mathbf L} d k
    \right\}
    \nonumber\\
    &=&
    \operatorname{exp}
    \left\{
        \left(
            \oint_{ \Gamma(\alpha) } -A_{12} d k
        \right)
        L_3
    \right\}.
    \label{eq_A_Wilson12}
\end{eqnarray}
The spin Wilczek-Zee connection which is a $\mathfrak{spin} \left( N \right)$-valued 1-form is then written as \cite{Wu2019}
\begin{equation}
    \bar{A} \left( k \right)
    = {\boldsymbol \beta}_{12} \left( k \right) \cdot {\mathbf t}
    = {\boldsymbol \beta}_{12} \left( k \right) \cdot \left( - \frac i 2 {\boldsymbol \sigma} \right),
    \label{eq_A_BWZ_2x2_12}
\end{equation}
where the components of ${\boldsymbol \sigma} = \left[ \sigma_1, \sigma_2, \sigma_3 \right]$ are the Pauli matrices
and ${\mathbf t} = \left(- i / 2 \right) {\boldsymbol \sigma}$.

From Eqs.~(\ref{eq_A_Wilson12}) and (\ref{eq_A_BWZ_2x2_12}), the topological charge \cite{Wu2019,Yang2020} is then expressed as
\begin{eqnarray}
    n_\Gamma
    &=&
    \operatorname{exp}
    \left\{
        \oint_{ \Gamma(\alpha) } \bar{A} \left( k \right) d k
    \right\}
    \nonumber\\
    &=&
    \operatorname{exp}
    \left\{
        \oint_{ \Gamma(\alpha) } {\boldsymbol \beta}_{12} \left( k \right) \cdot \left( - \frac i 2 {\boldsymbol \sigma} \right) d k
    \right\}
    \nonumber\\
    &=&
    \operatorname{exp}
    \left\{
        - \frac i 2 \sigma_3 \oint_{ \Gamma(\alpha) } -A_{12} d k.
    \right\}.
    \label{eq_A_n_Gamma_12}
\end{eqnarray}
If the integral $\oint_{ \Gamma(\alpha) } -A_{12} d k$ is $\pm \pi$, the charge $n_\Gamma$ becomes $\mp i\sigma_3$.


Now, if the closed loop encircles the nodal line formed by the bands $m=2$ and $n=3$, $\mathbf A \left( \mathbf k \right)$ in Eq.~(\ref{eq_A_BWZ_skewsymm}) is rewritten as $\mathbf A \left( k \right)={\boldsymbol \beta}_{23} \left( k \right) \cdot {\mathbf L} = -A_{23} L_1$ where ${\boldsymbol \beta}_{23} \left( k \right) = \left[ -A_{23}, 0,0 \right]$.
Substituting this into Eq.~(\ref{eqn:WilsonLoop}) gives
\begin{eqnarray}
    W
    &=&
    \operatorname{exp}
    \left\{
        \left(
            \oint_{ \Gamma(\alpha) } -A_{23} d k
        \right)
        L_1
    \right\}.
    \label{eq_A_Wilson23}
\end{eqnarray}
The spin Wilczek-Zee connection is $\bar{A} \left( k \right)= {\boldsymbol \beta}_{23} \left( k \right) \cdot {\mathbf t} = {\boldsymbol \beta}_{23} \left( k \right) \cdot \left( - \frac i 2 {\boldsymbol \sigma} \right)$, and the charge in Eq.~(\ref{eq_A_n_Gamma_12}) is then rewritten as
\begin{eqnarray}
    n_\Gamma
    &=&
    \operatorname{exp}
    \left\{
        - \frac i 2 \sigma_1 \oint_{ \Gamma(\alpha) }^{ } -\left[ A \left( k \right) \right]_{23} d k.
    \right\}.
    \label{eq_A_n_Gamma_23}
\end{eqnarray}
In the same way as the first case, we obtain $n_\Gamma = \mp i \sigma_1$ when $\oint_{ \Gamma(\alpha) }^{ } -A_{23} d k = \pm \pi$.

\subsection{Non-Abelian quaternion charges}
\label{section:non-Abelian}
To describe the non-Abelian band topology, Q. Wu et al. \cite{Wu2019} employed the quaternion numbers, ${\mathbb Q} = \left\{ \pm \boldsymbol i, \pm \boldsymbol j, \pm \boldsymbol k, \pm 1 \right\}$ (first written by the Irish mathematician, William Rowan Hamilton in 1843). 
The basis elements ${\boldsymbol i}$, ${\boldsymbol j}$, and ${\boldsymbol k}$ are defined such that ${\boldsymbol i}^2 = {\boldsymbol j}^2 = {\boldsymbol k}^2 = -1$. 
Their multiplication relations are ${\boldsymbol i}{\boldsymbol j} = {\boldsymbol k}$, ${\boldsymbol j}{\boldsymbol k} = {\boldsymbol i}$, and ${\boldsymbol k}{\boldsymbol i} = {\boldsymbol j}$. 
They all anticommute, that is, ${\boldsymbol i}{\boldsymbol j} = -{\boldsymbol j}{\boldsymbol i}$, ${\boldsymbol j}{\boldsymbol k} = -{\boldsymbol k}{\boldsymbol j}$, and ${\boldsymbol k}{\boldsymbol i} = -{\boldsymbol i}{\boldsymbol k}$. 
All these are summarized in Figure~\ref{fig:Schematic_Charge_IJK}c.
Interestingly, the Pauli matrices $\left\{ \mp i \sigma_1, \mp i \sigma_2, \mp i \sigma_3, \pm I \right\}$ exhibit the same properties as the quaternions and are isomorphic so that we can map ${\boldsymbol i} \mapsto -i \sigma_1$, ${\boldsymbol j} \mapsto -i \sigma_2$, ${\boldsymbol k} \mapsto -i \sigma_3$, and $1 \mapsto I$.
Due to the anticommutative properties of ${\boldsymbol i}$, ${\boldsymbol j}$, ${\boldsymbol k}$, or $-i \sigma_1$, $-i \sigma_2$, $-i \sigma_3$, their relations are regarded as non-Abelian.

Thus, the topological charges calculated in Eqs.~(\ref{eq_A_n_Gamma_12}) and (\ref{eq_A_n_Gamma_23}) can be regarded as the quaternions $\pm {\boldsymbol k}$ and $\pm {\boldsymbol i}$, respectively.

\subsection{Behavior of eigenstates of a 3$\times$3 Hamiltonian}
We apply the above formalism to a system expressed by a $3\times3$ Hamiltonian.
First, the rightmost side of Eq.~(\ref{eq_A_Wilson12}) is rewritten in rotation matrix form:
\begin{equation}
    W
    =
    \left[
        \begin{array}{ccc}
            \operatorname{cos} \left( \phi_{12} \right) & -\operatorname{sin} \left( \phi_{12} \right) & 0 \\
            \operatorname{sin} \left( \phi_{12} \right) & \operatorname{cos} \left( \phi_{12} \right) & 0 \\
            0 & 0 & 1
        \end{array}
    \right],
    \label{eq_A_WRot_12}
\end{equation}
where
\begin{equation}
    \phi_{12} = \oint_{ \Gamma(\alpha) } -A_{12} d k.
    \label{eq_A_phi_12}
\end{equation}
Let us suppose that $\left| u_{\mathbf k}^n \right\rangle$ ($n=1,2,3$) are the eigenstates of the Hamiltonian of the given system.
To satisfy Eqs.~(\ref{eq_A_BWZ_Mat_12}) and (\ref{eq_A_WRot_12}) we fix $\left| u_{\mathbf k}^3 \right\rangle$ as $\left[ 0,0,1 \right]$.
We also assume
$\left| u_{{\mathbf k} \left( \alpha \right)}^1 \right\rangle
=
\left[
    \operatorname{cos} \left( g \left( \alpha \right) \right),
    \operatorname{sin} \left( g \left( \alpha \right) \right),
    0
\right]$
and
$\left| u_{{\mathbf k} \left( \alpha \right)}^2 \right\rangle
=
\left[
    -\operatorname{sin} \left( g \left( \alpha \right) \right),
    \operatorname{cos} \left( g \left( \alpha \right) \right),
    0
\right]$
where $g \left( \alpha \right)$ is a real-valued arbitrary function that depends on the position of the closed-loop $\Gamma \left( \alpha \right)$ parametrized by $\alpha$.
For convenience, we set $g \left( 0 \right) = 0$.
The integral in Eq.~(\ref{eq_A_phi_12}) is written as
\begin{equation}
    \phi
    =
    \int_{ \alpha = 0 }^{ \alpha = 2\pi }
        - \left\langle u_{{\mathbf k} \left( \alpha \right)}^1
        \middle| \frac {\partial u_{{\mathbf k} \left( \alpha \right)}^2} {\partial \alpha}
          \right\rangle
        d \alpha
    = g \left( 2\pi \right)
    \label{eq_A_phi_12_2pi}
\end{equation}
In the above, we mentioned that the quaternion $\mp i \sigma_3$ is obtained if this integral is $\pm \pi$.
Thus, we can deduce $g \left( 2\pi \right) = \pm \pi$, and one can define $g \left( \alpha \right) = \pm \alpha / 2$ \cite{Wu2019,Yang2020}.
This means that we can plot the eigenstates $\left| u_{\mathbf k}^1 \right\rangle$ and $\left| u_{\mathbf k}^2 \right\rangle$ rotating around the fixed $\left| u_{\mathbf k}^3 \right\rangle$ by $\pm \pi$.

In the same manner, the rightmost side of Eq.~(\ref{eq_A_Wilson23}) becomes
\begin{equation}
    W
    =
    \left[
        \begin{array}{ccc}
            1 & 0 & 0 \\
            0 & \operatorname{cos} \left( \phi_{23} \right) & -\operatorname{sin} \left( \phi_{23} \right) \\
            0 & \operatorname{sin} \left( \phi_{23} \right) & \operatorname{cos} \left( \phi_{23} \right) \\
        \end{array}
    \right],
    \label{eq_A_WRot_23}
\end{equation}
where
\begin{equation}
    \phi_{23} = \oint_{ \Gamma(\alpha) }^{ } -A_{23} d k.
    \label{eq_A_phi_23}
\end{equation}
We assume $\left| u_{\mathbf k}^1 \right\rangle = \left[ 1,0,0 \right]$,
$\left| u_{{\mathbf k} \left( \alpha \right)}^2 \right\rangle
=
\left[
    0,
    \operatorname{cos} \left( g \left( \alpha \right) \right),
    \operatorname{sin} \left( g \left( \alpha \right) \right)
\right]$
, and
$\left| u_{{\mathbf k} \left( \alpha \right)}^3 \right\rangle
=
\left[
    0,
    -\operatorname{sin} \left( g \left( \alpha \right) \right),
    \operatorname{cos} \left( g \left( \alpha \right) \right)
\right]$
with the same $g \left( \alpha \right)$. The integral in Eq.~(\ref{eq_A_phi_23}) becomes
\begin{equation}
    \phi
    =
    \int_{ \alpha = 0 }^{ \alpha = 2\pi }
        - \left\langle u_{{\mathbf k} \left( \alpha \right)}^2
        \middle| \frac {\partial u_{{\mathbf k} \left( \alpha \right)}^3} {\partial \alpha}
          \right\rangle
        d \alpha
    = g \left( 2\pi \right)
    \label{eq_A_phi_23_2pi}
\end{equation}
and we get similar results; the eigenstates $\left| u_{\mathbf k}^2 \right\rangle$ and $\left| u_{\mathbf k}^3 \right\rangle$ rotate by $\pm \pi$ around $\left| u_{\mathbf k}^1 \right\rangle$ that corresponds to the quaternion $\mp i \sigma_1$.

Thus, if the nodal line system is described by a 3$\times$3 Hamiltonian, the eigenstates ${\mathbf u}_{\mathbf k}^1$, ${\mathbf u}_{\mathbf k}^2$, ${\mathbf u}_{\mathbf k}^3$ along the closed loop can be calculated and plotted along an arbitrary coordinate system (see Figure~\ref{fig:Schematic_Charge_IJK}b). After collecting the eigenstates at the origin, the rotation behavior of the eigenstates indicates the corresponding topological charge. For example, in Figure~\ref{fig:Schematic_Charge_IJK}d, ${\mathbf u}_{\mathbf k}^3$ is fixed while ${\mathbf u}_{\mathbf k}^1$ and ${\mathbf u}_{\mathbf k}^2$ show the $+\pi$-rotation. Then, this is considered as the quaternion charge ${\boldsymbol k}$. Figure~\ref{fig:Schematic_Charge_IJK}e and f also show similar behaviors, thereby their charges are ${\boldsymbol i}$ and ${\boldsymbol j}$, respectively.

\subsection{Correlations for full-vector field problems}
Let us think more about the rotation behaviors of the eigenstates $\left| u_{\mathbf k}^1 \right\rangle$, $\left| u_{\mathbf k}^2 \right\rangle$, and $\left| u_{\mathbf k}^3 \right\rangle$.
If we denote the starting point of the closed loop as ${\mathbf k}_0$ and choose the orthonormal coordinates placed along $\left| u_{{\mathbf k}_0}^1 \right\rangle$, $\left| u_{{\mathbf k}_0}^2 \right\rangle$, and $\left| u_{{\mathbf k}_0}^3 \right\rangle$, these eigenstates $\left| u_{{\mathbf k}_0}^n \right\rangle$ are the same as the unit vectors of the coordinate system. For an arbitrary orthonormal coordinate system $\left| e^n \right\rangle$, the eigenstates $\left| u_{\mathbf k}^n \right\rangle$ can be mapped to $\left| {\left( u' \right)}_{\mathbf k}^n \right\rangle$ by the following rotation matrix
\begin{equation}
    {\mathbf R} = \sum_{n=1}^3 \left| e^n \right\rangle \left\langle u_{{\mathbf k}_0}^n \right|.
\end{equation}
The resulting new eigenstates $\left| {\left( u' \right)}_{\mathbf k}^n \right\rangle$ are explicitly written as
\begin{equation}
    \left| {\left( u' \right)}_{\mathbf k}^n \right\rangle
    =
    \left[
        \left\langle
        u_{{\mathbf k}_0}^1
        |
        u_{\mathbf k}^n
        \right\rangle
        ,
        \left\langle
        u_{{\mathbf k}_0}^2
        |
        u_{\mathbf k}^n
        \right\rangle
        ,
        \left\langle
        u_{{\mathbf k}_0}^3
        |
        u_{\mathbf k}^n
        \right\rangle
    \right] ^T .
    \label{eq_A_3x3_eigenstate_in_newcoord}
\end{equation}
Now, we want to render a plot similar to Figure~\ref{fig:Schematic_Charge_IJK}d-f when the system is described not only by a 3$\times$3 matrix but also a geometry-dependent Hamiltonian, e.g., as in photonic crystals. In this case, an eigenstate $\left| \psi_{\mathbf k}^n \right\rangle$ is a function of the three-dimensional position vector \cite{beekman_2017}. Similar to Eq.~(\ref{eq_A_3x3_eigenstate_in_newcoord}) the following correlations can be defined \cite{Park_2021a_ACSPhotonics}:
\begin{equation}
    {\mathbf C}_{\mathbf k}^n
    =
    \left[
        \left\langle
        \psi_{{\mathbf k}_0}^1
        |
        \psi_{\mathbf k}^n
        \right\rangle
        ,
        \left\langle
        \psi_{{\mathbf k}_0}^2
        |
        \psi_{\mathbf k}^n
        \right\rangle
        ,
        \left\langle
        \psi_{{\mathbf k}_0}^3
        |
        \psi_{\mathbf k}^n
        \right\rangle
    \right].
    \label{eq_A_Correlation}
\end{equation}
In the same manner, if the correlations ${\mathbf C}_{\mathbf k}^1$ and ${\mathbf C}_{\mathbf k}^2$ rotate by $\pm \pi$ around ${\mathbf C}_{\mathbf k}^3$, it corresponds to the quaternion $\mp i \sigma_3$.
And if the correlations ${\mathbf C}_{\mathbf k}^2$ and ${\mathbf C}_{\mathbf k}^3$ rotate by $\pm \pi$ around ${\mathbf C}_{\mathbf k}^1$, it corresponds to the quaternion $\mp i \sigma_1$.

\begin{figure}
    \centering
    \includegraphics{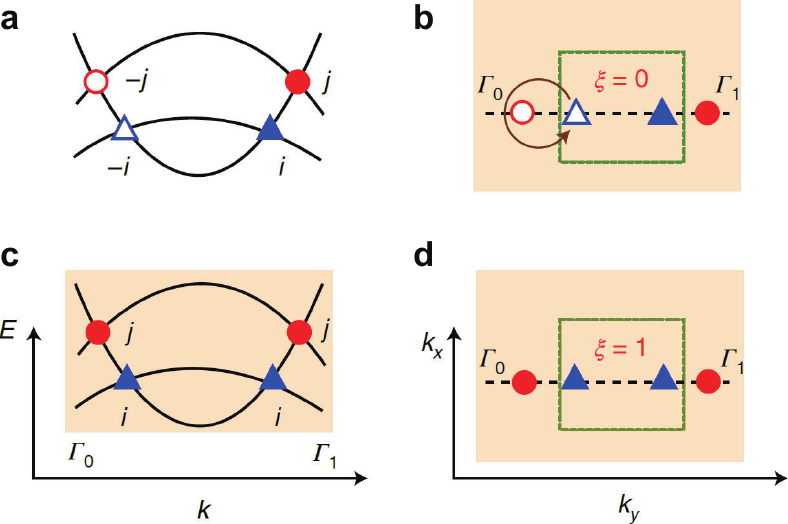}
    \caption{
        \label{fig:PatchEulerClass}    
            Schematic of a three-band system exhibiting the non-Abelian band nodes in two-dimensional momentum space. (a) Band structure with oppositely charges nodes. The blue triangles and red circles indicate the nodes whose charges are $\pm \boldsymbol i$ and $\pm \boldsymbol j$, respectively. The open and filled symbols correspond to the plus and minus signs, respectively. (b) Braiding of the open triangle node around the open circle node. (c),(d) Band structure with the nodes whose charges changed by a factor of $-1$ after the tuning of (b). The patch Euler class is now $1$. Reproduced with permission from Ref.~\cite{Jiang2021}. Copyright 2021, Nature Portfolio (a-d).
    }
\end{figure}
\subsection{Evolution of degeneracies and the patch Euler class}
\label{section:patchEulerClass}
In the previous sections, the non-Abelian topological charges in three-band systems were introduced. On top of discovering the nodal line systems with non-Abelian charges, it has been studied how the non-Abelian charged degeneracies evolve with the tuning of the Hamiltonian. Such evolutions include the annihilation or creation of the degeneracies \cite{Bzdusek2017,Bouhon2019,Sun_PRL_2018,Bouhon_NatPhys_2020,Peng2021}. 
To understand whether the degeneracies are annihilated or not, the patch Euler class was introduced in Ref. \cite{Bouhon2020,Bouhon_NatPhys_2020}.

In a recent experimental work \cite{Jiang2021}, B. Jiang et al. demonstrated the evolution of degeneracies with non-Abelian charges and patch Euler class. The experimentally observed evolution of degeneracies clearly shows the creation and annihilation of degeneracies although they considered Dirac nodes in two-dimensional momentum space instead of the nodal lines in three-dimensional momentum space. For the three-band system in two-dimensional momentum space illustrated in Figure~\ref{fig:PatchEulerClass}a, the nodes between the first and second (second and third) bands have the charges $\pm \boldsymbol i$ ($\pm \boldsymbol j$). Here, the two charges $\pm \boldsymbol i$ ($\pm \boldsymbol j$) have opposite signs. Thus, the oppositely charged $\pm \boldsymbol i$ (or $\pm \boldsymbol j$) can be annihilated pairwise and the patch Euler class is zero. However, braiding the open triangle around the open circle (see Figure~\ref{fig:PatchEulerClass}b) makes these two flip their charges to positive as shown in Figure~\ref{fig:PatchEulerClass}(c). As a result, the patch Euler class becomes one (see Figure~\ref{fig:PatchEulerClass}d). A patch Euler class of $\pm 1$ means that when the two nodes merge together, they form a stable quadratic node with frame charge $q=-1$ rather than undergoing pairwise annihilation.

\section{Topological nodal lines in photonic systems}
\label{section:PhotonicExample}
\begin{figure*}[ht]
    \centering
    \includegraphics{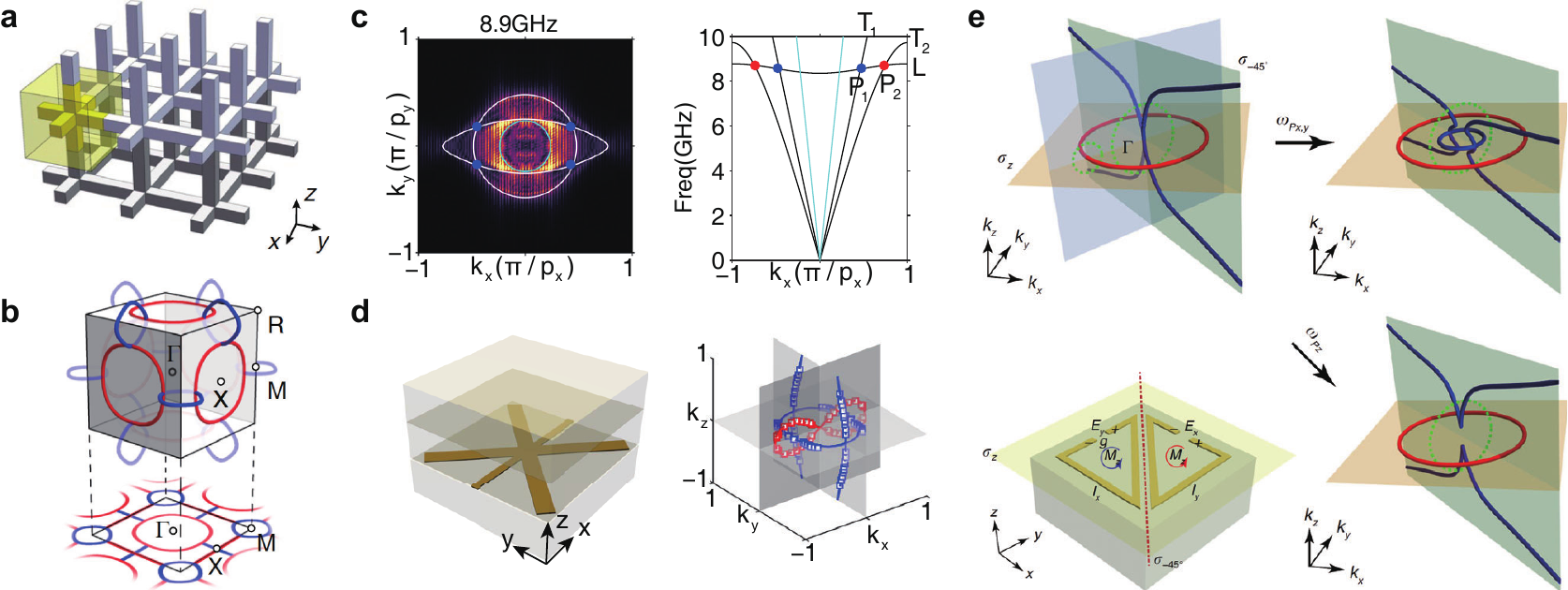}
    \caption{
        \label{fig:NodalLine_Metamaterials}
 Examples of nodal line topological metamaterials. (a) Schematic of the metallic mesh structure for generating a nodal chain. (b) The illustration of the nodal chain consists of two types of nodal lines (blue and red) in the cubic Brillouin zone. (c) Equifrequency contour and band structure of the non-Abelian nodal line metamaterial reminiscent of a biaxial crystal at low frequencies (d) Schematic of the unit cell and the nodal lines in the Brillouin zone. (e) The transitions between orthogonal nodal chain, in-plane nodal chain, and separated nodal lines are observed in the illustrated bianisotropic metamaterial. Reproduced with permission from Ref.~\cite{Yan_2018_NatPhys}. Copyright 2018, Nature Portfolio (a-b); Reproduced with permission from Ref.~\cite{Yang2020}. Copyright 2020, American Physical Society (c-d); Reproduced with permission from Ref.~\cite{Wang_LSA_2021}. Copyright 2021, Nature Portfolio (e).}
\end{figure*}
As we described in Section~\ref{section:non-Abelian}, nodal line degeneracies can be described by the Weyl Hamiltonian without $\sigma_2$ term. Therefore, the goal of designing a structure with nodal lines is to find geometrical or coupling parameters for a structure that has such a dispersion. 
Although there is no general recipe that can be applied to different systems, the spatial symmetry consideration can be a good guide to find nodal lines because a structure with nodal lines respects $\mathcal{P}$ and $\mathcal{T}$ symmetry as mentioned earlier. For instance, if we know a structure with Weyl degeneracy, that respects $\mathcal{P}$ or $\mathcal{T}$ symmetry only, we can start from the structure to recover both symmetries. Alternatively, if we  start with a structure with excessive symmetries including  $\mathcal{P}$ and $\mathcal{T}$ symmetries, which is the case for most of the Bravais lattices, we need to introduce perturbations in the direction of reducing the number of symmetries. 
Numerical simulations are often used to see how the dimension of degeneracies changes during the symmetry reduction process.

In this section, we will introduce examples of nodal lines in photonic systems using metamaterials, metallic photonic crystals and dielectric photonic crystals. 

\subsection{Metamaterials and metallic photonic crystals}
\label{section:metamaterial}

Photonic metamaterials emerge as a prominent light matter interaction platform and have attracted enormous research interest within the past two decades \cite{simovski_tretyakov_2020,Soukoulis_NatPhoton_2011,Kadic_NatRevPhys_2019,Jahani_NatNanotech_2016}. Utilizing the quasi-homogeneity and by manipulating the constituent deep-subwavelength units, so-called meta-atoms, they can create collective responses to photons far beyond the scope of natural materials, for instance, negative refraction \cite{Pendry_PRL_2000,Hoffman_NatMat_2007}, strong anisotropy \cite{Luo_PRL_2014,Fang_PRB_2009}, hyperbolicity \cite{Yao_Science_2008,Poddubny_NatPhoton_2013}, strong optical activity \cite{Pendry_Science_2004,SZhang_PRL_2009}, etc. 
Metamaterials have provided a reliable and convenient guideline towards artificial photonic materials, assisted by the rich design experience accumulated in the past two decades. Providing new degrees of freedom in photonic material design, researchers have found rich fundamentally new physics and applications with metamaterials. 

Topological photonic metamaterials have emerged in recent years as a salient topic within the grand regime of topological photonics. 
For instance, bianisotropic metamaterials were used to realize a photonic topological insulator \cite{Khanikaev_NatMat_2013}. In fact, topological phenomena in continuous photonic medium have a long-standing history including the renowned Pancharatnam-Berry phase in polarization space \cite{Pancharatnam_PIAS_1956}, and the conical diffraction in biaxial crystals that is a direct consequence of the quantized Berry phase of the Dirac point \cite{Turpin_lpor_2016}.

The first topological metamaterial was designed in 2015 \cite{Gao_PRL_2015}, in which a composite response from hyperbolicity and chirality introduces Fermi surfaces with distinct Chern numbers that uni-directional surface state connects. 
Various topological semimetals have also been discovered in metamaterials, including Weyl nodes \cite{Gao_NatComm_2016} and Dirac nodes \cite{Guo_PRL_2017}. 
Naturally, metamaterials, or equivalently effective medium methods, play an important role in the construction of topological nodal lines. 

An ideal photonic nodal line was discovered in a type-I hyperbolic metamaterial \cite{Gao2018}. Here `ideal' refers to that the nodal line is free from coexisting trivial modes in the bulk. The band crossing happens between effective longitudinal and transverse modes in which interactions are eliminated due to their mismatched field polarizations. Despite being an accidental degeneracy, the band crossing is imposed by the engineered nonlocal response in the metamaterial exerted by the glide reflection symmetry. 
In terms of crystallographic symmetry, these nodal lines are protected by mirror symmetry and by introducing mirror symmetry breaking terms, for instance bianisotropy, these nodal lines are instantly gapped and give rise to vortex-like distributed Berry curvatures \cite{Yang_2021_NatComm}. 

An ideal type-II nodal line has been discovered recently in Bragg reflection mirror type layered photonic crystals as the phase transition point between trivial and non-trivial Zak phase regimes \cite{Deng_2021_arXiv}.
Definition of the ‘type-II’ follows the classification of Weyl points \cite{Soluyanov2015}, meaning the highly tilted contact between bands. It exhibits a ring-like contact between electron and hole pocket, distinguished from the donut-like Fermi surface in type-I nodal line semimetals \cite{Gao2018}.
The nodal chain was experimentally introduced using a three-dimensional metallic-mesh structure (Figure~\ref{fig:NodalLine_Metamaterials}a) in microwave scale \cite{Yan_2018_NatPhys}, which was the original design of a metallic metamaterial with extremely low plasma frequency \cite{Pendry_PRL_1996}, though the nodal chain was found far above the plasma frequency and cannot be explained by effective medium theory. 
The nodal chain in Figure~\ref{fig:NodalLine_Metamaterials}b, although it consists of two colored nodal rings, is formed by the same adjacent two bands. 
This study also examined the drumhead surface states, a sheet of surface dispersion enclosed by the projected nodal line bulk states on the surface Brillouin zone. 

Another example of crystallographic symmetry mediated optical response of meta-atoms was demonstrated in the discovery of hourglass nodal lines in a photonic metamaterial \cite{Lingbo_PRL_2019}. 
Although it may seem that band topologies can be solely determined by global crystallographic symmetries, the interplay between them and local optical responses is surprisingly rich in new physics, for instance, the hidden symmetries that are unforeseen by crystallographic group theory \cite{Xiong_LSA_2020}.
It is, however, worth noting that without exquisite design, the touching point between equi-frequency contours in natural biaxial crystal forms a 3-dimensional nodal chain if the bands at higher momentum are considered cut off and flattened by the Brillouin zone boundary (Figure~\ref{fig:NodalLine_Metamaterials}c). Utilizing this property and the extreme anisotropy provided by metamaterials, researchers have constructed and measured nodal-link metamaterials in the microwave regime \cite{Yang2020}. 
The nodal link in this metamaterial is formed by the lowest three bands (see Figure~\ref{fig:NodalLine_Metamaterials}c, d). The surface bound states in the continuum are another achievement of this study. 
Moreover, in a metamaterial with explicitly broken inversion symmetry through bianisotropic optical activities (Figure~\ref{fig:NodalLine_Metamaterials}e), transition between different types of nodal chains is observed by engineering the optical resonances of meta-atoms \cite{Wang_LSA_2021}. One remarkable aspect of these studies is that the topological nature of the nodal line was well described by the non-Abelian band topology \cite{Wu2019}, and can be conveniently derived from the effective Hamiltonian model stemming from the effective medium theory without referring to microscopic electromagnetic fields within the structures.

\begin{figure*}[ht]
    \centering
    \includegraphics{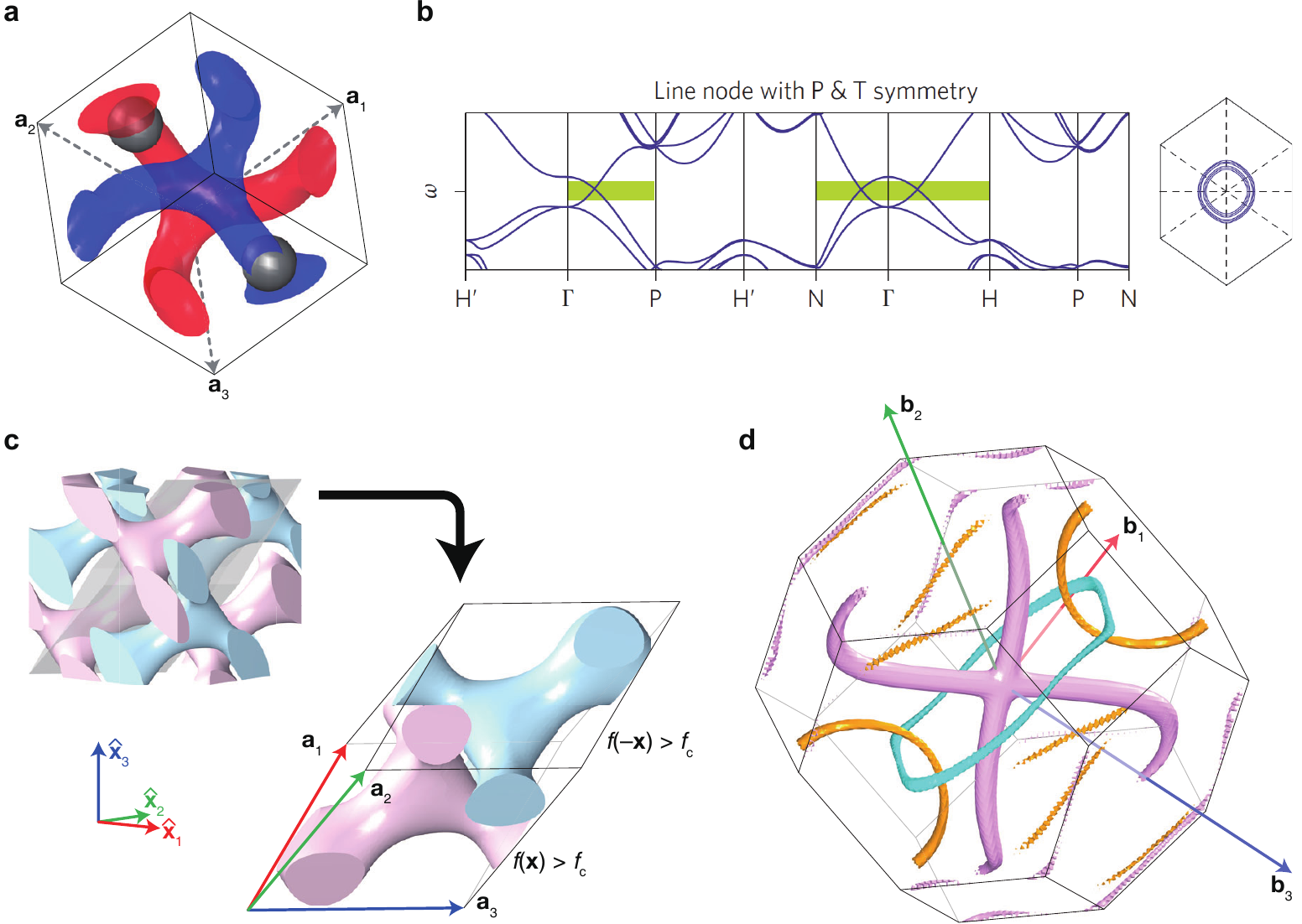}
    \caption{
        \label{fig:Dielectric_NodalLine}
            Examples of nodal lines in dielectric photonic crystals. (a) Inversion symmetric double gyroid structure that has air spheres. (b) Band structure of (a) that exhibits a nodal ring in the momentum space. (c) Inversion symmetric and anisotropic double diamond structure. (d) Nodal link, nodal chain, and nodal lines in momentum space of the structure in (c). Reproduced with permission from Ref.~\cite{Lu2013}. Copyright 2013, Nature Portfolio (b); Reproduced with permission from Ref.~\cite{Park_2021a_ACSPhotonics}. Copyright 2021, American Chemical Society (c-d).
    }
\end{figure*}
\subsection{Dielectric photonic crystals}
\label{section:photonic}

In this section, we explain how to realize nodal lines in dielectric materials with two examples: double gyroid and double diamond structures. Although fabricating dielectric photonic crystals is challenging, these crystals have advantages such as scalability and convenience in theoretical descriptions.

The nodal ring was theoretically realized using an $\mathcal{P}$-symmetric double gyroid \cite{Lu2013}.
The well-known single gyroid in $O$-symmetry is defined by a set $\mathbf x = \left[ x_1 , x_2 , x_3 \right]$ such that
\begin{eqnarray}
    g \left( \mathbf x \right)
    &=&
    \operatorname{sin} \left( X_1 \right) \operatorname{cos} \left( X_2 \right)
    + \operatorname{sin} \left( X_2 \right) \operatorname{cos} \left( X_3 \right)
    \nonumber\\
    &+& \operatorname{sin} \left( X_3 \right) \operatorname{cos} \left( X_1 \right) > g_c > 0 ,
\end{eqnarray}
and its space group is $I 4_1 32$ (No. 214) \cite{Fruchart_PNAS_2018,Park_2020_ACSPhotonics}. Here, $X_i = \left( 2\pi / a \right)x_i$ is a local coordinate where $a$ is a lattice constant. 
In the study in Ref.~\cite{Lu2013}, the reduced-symmetric single gyroid is created by introducing an air sphere of radius $0.13a$ located at $\left[ 1/4, -1/8, 1/2 \right]a$ in the single gyroid. Then, a double gyroid was created by combining this single gyroid and its counterpart while satisfying inversion symmetry, as shown in Figure~\ref{fig:Dielectric_NodalLine}a. The photonic band structure reveals that the set of degeneracies between the forth and fifth bands form the nodal ring, as shown in Figure~\ref{fig:Dielectric_NodalLine}b.

Very recently, a double diamond photonic crystal exhibiting nodal link, nodal chain, and nodal lines all at once was reported \cite{Park_2021a_ACSPhotonics}. When the lattice vectors are given by ${\mathbf a}_1 = a/2 \left[ 0,1,1 \right]$, ${\mathbf a}_2 = a/2 \left[ 1,0,1 \right]$, and ${\mathbf a}_1 = a/2 \left[ 1,1,0 \right]$ with the lattice constant $a$, the double diamond is defined by a set $\mathbf x$ satisfying $f \left( \pm \mathbf x \right) > f_c > 0 $. Each inequality with plus or minus sign corresponds to each single diamond, so that the two single diamonds are inversion symmetric. Here, the function $f \left( \mathbf x \right)$ is given by
\begin{eqnarray}
    f \left( \mathbf x \right)
    &=&
    A_0 \operatorname{sin} \left( X_1 + X_2 + X_3 \right)
    \nonumber\\
    &+& \sum_{i=1} ^3 A_i \operatorname{sin} \left( X_1 + X_2 + X_3 - 2 X_i \right)
\end{eqnarray}
where $\mathbf X = \left[ X_1 , X_2 , X_3 \right] = \left( 2 \pi / a \right) \left( \mathbf x -\gamma /2 \sum_{i=1} ^3 {\mathbf a}_i \right)$ is a local coordinate that expresses the translation of the single diamonds along the $\pm \left[ 1,1,1 \right]$-directions, adjusted by the coefficient $\gamma$. Selecting $A_0 = A_1 = A_2 = A_3$ and $\gamma = 0$ generates the conventional diamond structure \cite{Wohlgemuth_Macromelecules_2001,Angelova_JInorgOrgPolMat_2015,Barriga_SoftMatter_2015,La_NatComm_2018,Sheng_NanoRes_2020}. However, this study selected different coefficients $A_i$ and non-zero $\gamma$ (see Figure~\ref{fig:Dielectric_NodalLine}c) to destroy as many symmetries as possible. The resulting structure is anisotropic so that the double diamond has only inversion and translational symmetries.

In momentum space the degeneracies between the first and second bands form the nodal chain, the degeneracies between the third and fourth bands form the simple nodal lines, and the degeneracies between the third, fourth, and fifth bands form the nodal link (see Figure~\ref{fig:Dielectric_NodalLine}d).
The nodal chain is centered at the $\Gamma$-point. Two pink nodal lines (between the first and second bands) depart the $\Gamma$-point and arrive on a single boundary. Another two pink nodal lines are centrosymmetric to the first two pink nodal lines. The two boundaries that these nodal lines touch are parallel, and their normal vectors are commonly along the ${\mathbf b}_2$-direction. Thus, the nodal chain is infinitely connected along the ${\mathbf b}_2$-direction.

Meanwhile, the nodal link consists of the non-touching orange (between the third and fourth bands) and cyan rings (between the fourth and fifth bands), as shown in Figure~\ref{fig:Dielectric_NodalLine}d. The cyan ring is centered at the $\Gamma$-point, and the center of the orange ring is located on a boundary. Due to the periodicity of the first Brillouin zone, this link is infinitely connected.

One more significance of this research \cite{Park_2021a_ACSPhotonics} is that the correlation vectors in Eq.~(\ref{eq_A_Correlation}) were first introduced. By employing the correlation vectors, the non-Abelian topological nature of the nodal link could be directly calculated from the full-vector field eigenstates, instead of using the $3\times3$ effective Hamiltonian, thus the topological charges ${\mathbb Q} = \left\{ \pm \boldsymbol i, \pm \boldsymbol j, \pm \boldsymbol k, \pm 1 \right\}$ were completely deduced from the numerically calculated nodal link. 

A simple post-processing calibration was also proposed in this photonic study \cite{Park_2021a_ACSPhotonics}. To get the topological charge $\pm \boldsymbol j$, a closed loop that encloses two nodal lines exhibiting the topological charges $\pm \boldsymbol k$ and $\pm \boldsymbol i$ should be considered. Although we illustrate the quaternion charge $\pm \boldsymbol j$ in Figure~\ref{fig:Schematic_Charge_IJK}f, the eigenstates or correlations calculated by Figure~\ref{fig:Schematic_Charge_IJK}a and b do not generally show results as in Figure~\ref{fig:Schematic_Charge_IJK}f. In many cases, they exhibit the $\pi$-disclinations of ${\mathbf u}_{\mathbf k} ^3$ and ${\mathbf u}_{\mathbf k} ^1$ at ${\mathbf k}_0$ of the loop, implying the topological charge $\pm \boldsymbol j$. Here, the $\pi$-disclinations mean that, before and after winding along the loop, the directions of ${\mathbf u}_{\mathbf k} ^3$ and ${\mathbf u}_{\mathbf k} ^1$ are rotated by $\pi$. To observe the rotation behaviors of the eigenstates or correlations more clearly, this study \cite{Park_2021a_ACSPhotonics} introduces a post-processing calibration method. First, at a point $\mathbf k$, a rotation matrix $\mathbf R \left( \mathbf k \right)$ is defined so that it maps ${\mathbf u}_{\mathbf k} ^2$ to ${\mathbf u}_{{\mathbf k}_0} ^2$ around $\mathbf r \left( \mathbf k \right) = {\mathbf u}_{\mathbf k}^2 \times {\mathbf u}_{{\mathbf k}_0} ^2$. Then, ${\mathbf u}_{\mathbf k} ^1$ and ${\mathbf u}_{\mathbf k} ^3$ are also rotated by $\mathbf R \left( \mathbf k \right)$. This process is done for all $\mathbf k$ on the closed loop, to get a result like Figure~\ref{fig:Schematic_Charge_IJK}f.

\subsection{Optical frequency synthetic dimension}
\label{section:synthetic}
As described in the previous sections, diverse topologies of nodal lines can be created using metamaterials and photonic crystals because one can design a periodic structure and tune geometric and optical parameters. However, the freedom in photonic designs is not simply limited to a spatially periodic system but one can extend the design freedom into the frequency domain based on a new concept called `synthetic dimension'. 
Indeed, this can provide another possibility to create more complex topologies. 
Very recently, K. Wang et al. \cite{Wang2021} showed that topologies such as unknots, Hopf links and trefoils can be created using two coupled ring resonators and a dielectric waveguide by modulating the phase and amplitude in one of the ring resonators. 
In their work, the frequency synthetic dimension is created by multiple resonance frequencies of the unperturbed ring resonators which are periodic in frequency space and the complex energy spectra are obtained by measuring the detuning of the resonance frequencies. 
By doing so, the wavevector-energy space $(k, \mathrm{Re}(E), \mathrm{Im}(E))$ becomes the parameter space where the optical bands with different topologies can exist. 
It is worth noting that the topology of two bands is considered instead of nodal lines originating from two different bands. 

\section{Topological nodal lines in other systems}
\label{section:example}

In this section, we review recent achievement in finding nodal lines in electronic crystals, phononic crystals and electrical circuits. 
Historically, the study on nodal lines in electronic crystals started earlier than all other systems.  
Several review papers have been published on nodal line semimetals \cite{Fang2016, Gao2019}.
Therefore, in our review, we highlight a few important achievements in electronic crystals.
\begin{figure}
    \centering
    \includegraphics{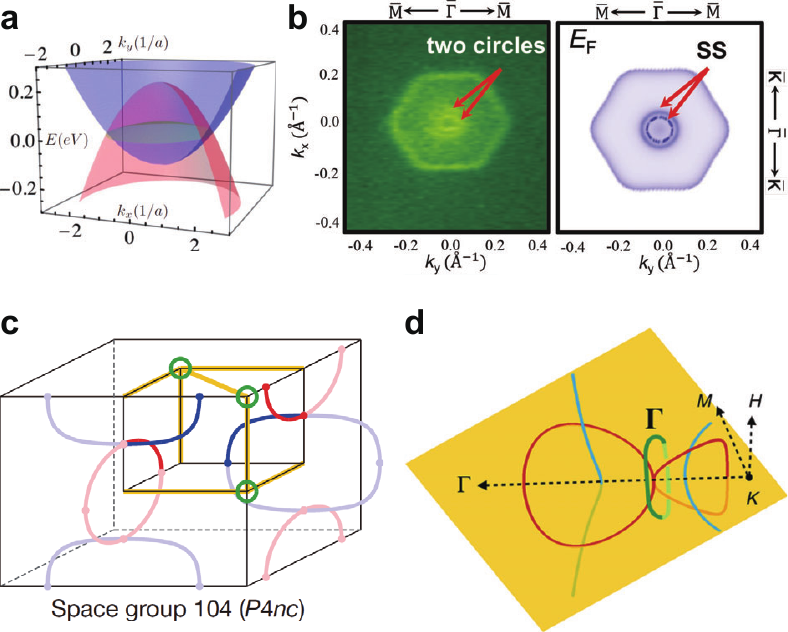}
    \caption{
        \label{fig:ElectronicCrystals}    
            Nodal lines and nodal chains in electronic crystals. 
            (a) Drumhead surface states (green) bounded by the nodal line in Ca$_3$P$_2$.
            (b) Surface states in Mg$_3$Bi$_3$ measured by ARPES.
            (c) Nodal chains.
            (d) nodal lines in IrF$_4$.
            Reproduced with permission from Ref.~\cite{Chan_PRB_2016}. Copyright 2021, American Physical Society (a); Reproduced with permission from Ref.~\cite{Chang_AdvSci_2019}. Copyright 2019, WILEY-VCH Verlag GmbH \& Co. KGaA, Weinheim (b); Reproduced with permission from Ref.~\cite{Bzdusek2016}. Copyright 2016, Nature Portfolio (c); Reproduced with permission from Ref.~\cite{Wu2019}. Copyright 2019, American Association for the Advancement of Science (d).
    }
\end{figure}
\begin{figure*}[ht]
    \centering
    \includegraphics{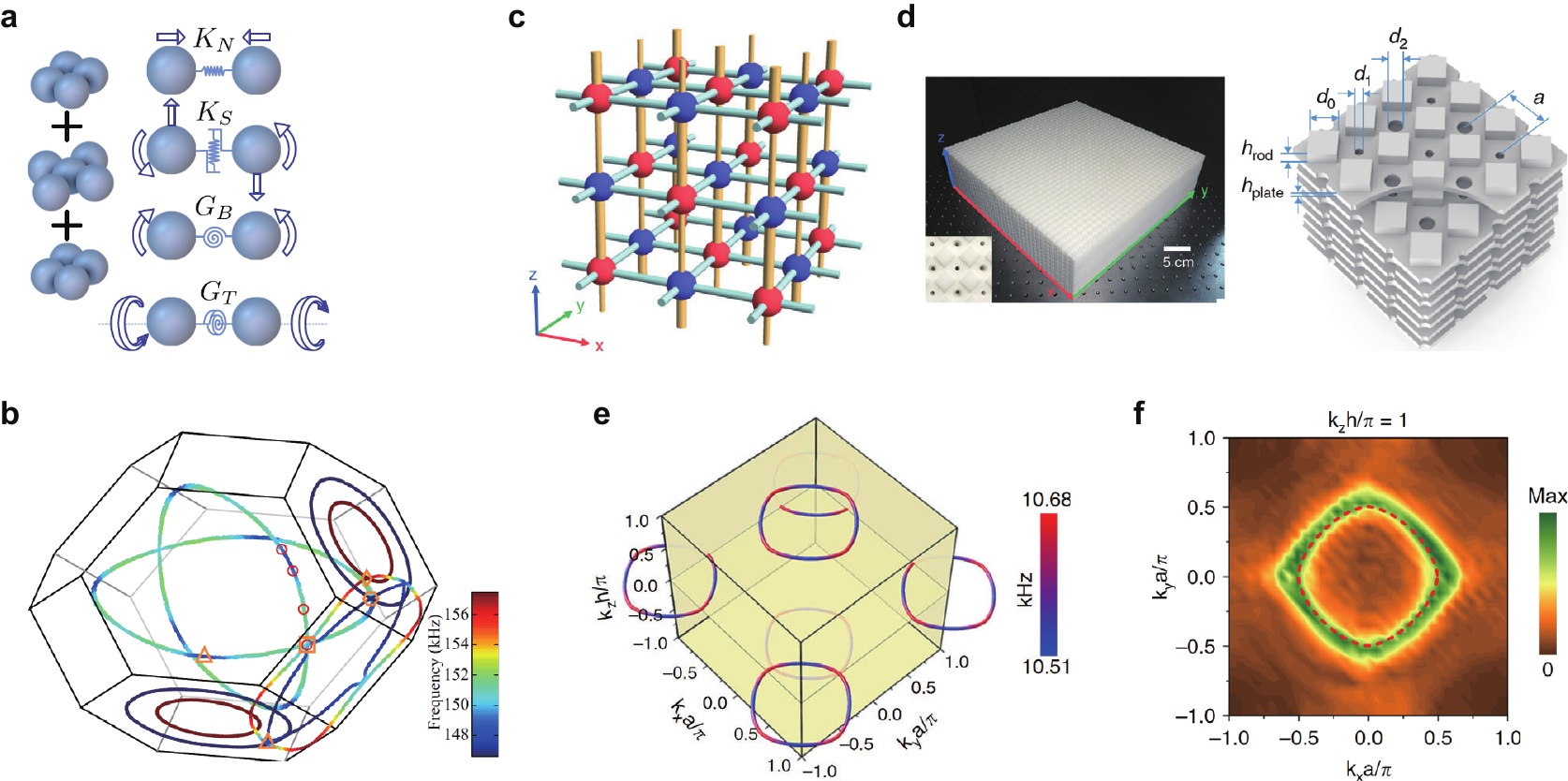}
    \caption{
        \label{fig:PnononNodalRing}    
            Nodal lines in phononic crystals. (a) Schematic of a granular metamaterial toward phononic nodal chain. Four types of interactions between two beads in the nearest-neighbors are also illustrated. (b). Simulation results showing the nodal chain in momentum space. (c) An illustration of the tight-binding Hamiltonian in Eq.~(\ref{eq_Phonon_Hamiltonian}) with different values of $t$ and $\delta t$. (d) Photograph (left) and schematics (right) of a phononic crystal exhibiting nodal rings. (e-f) Simulation and experimental results showing nodal rings, respectively. Reproduced with permission from Ref.~\cite{Merkel_CommPhys_2019}. Copyright 2019, Nature Portfolio (c-f); Reproduced with permission from Ref.~\cite{Deng2019}. Copyright 2019, Nature Portfolio (c-f).
    }
\end{figure*}
\subsection{Electronic crystals}
\label{section:electronic}
The discovery of nodal lines starts from the prediction of the cubic
antiperovskite material Cu$_{3}$NX, where X=\{Ni; Cu; Pd; Ag; Cd\}, as ${\mathbb Z}_2$ protected topological semimetals when ignoring spin-orbit interaction \cite{Rui_Yu_PRL_2015,Kim_PRL_2015}. 
This material holds one-dimensional Dirac line nodes and two-dimensional nearly-flat surface states, protected by $\mathcal{P}$ and $\mathcal{T}$ symmetries. 
In particular, the 2D surface states are bounded by the projected Dirac line nodes and because of this, they were called drumhead states (Figure \ref{fig:ElectronicCrystals}a) in the field afterwards. 
In the same work by Y. Kim et al. \cite{Kim_PRL_2015}, they showed that nearly flat surface states exist in Cu$_{3}$NX. 
Additionally, Y. H. Chan et al. showed that the drumhead surface states of Ca$_3$P$_2$ exist due to a quantized Berry phase and the ${\mathbb Z}_2$ topological invariants were defined similarly as in strong topological insulators \cite{Chan_PRB_2016}.
It is worth noting that, before the surge of the search for the drumhead edge states, a topologically protected flat band has drawn  attention because it can promote surface superconductivity with an infinite density of states \cite{Kopnin2011}.
Recently, the drumhead surface states were also shown in phononic crystals \cite{Deng2019}. Y. Wang et al. analyzed the flatness and boundedness of photonic drumhead surface states using a simple cubic lattice of metals \cite{Wang_SciRep_2021}. 

Soon after, other nodal line materials in the absence of spin-orbital interaction were predicted, such as alkaline-earth compounds AX$_{2}$ (A = Ca, Sr, Ba; X = Si, Ge, Sn) \cite{Huang_PRB_2016}, the CaP$_3$ family of materials \cite{Xu_PRB_2017,Takane_PRB_2018}, some of which are experimentally demonstrated, such as CaCdSn \cite{Laha_PRB_2020} and Mg$_3$Bi$_3$ \cite{Chang_AdvSci_2019} (Figure \ref{fig:ElectronicCrystals}b). 
Furthermore, topological semimetals can be classified into two types according to the tilting degree of the fermion cone. Type-II nodal lines in CaPd \cite{Liu_PRB_2018} and Mg$_3$Bi$_3$ \cite{Chang_AdvSci_2019} were proposed. Finally, topological spinful nodal lines were found to exist in TlTaSe$_2$ \cite{Bian_PRB_2016}, ZrSiSe and ZrSiTe \cite{Jin_Hu_PRL_2016} by including spin-orbit interaction as well.

More recently, diverse topologies have been reported in electronic crystals.  For example, nodal chains were predicted in IrF$_4$ (Figure \ref{fig:ElectronicCrystals}c) \cite{Bzdusek2017}  and nodal links were shown in Sc \cite{Wu2019}. Interestingly, non-Abelian topological invariants (see Subsection~\ref{section:non-Abelian}) were shown to exist in those multiple nodal line structures and this work has inspired much additional research on non-Abelian topology including the one in photonic crystals \cite{Park_2020_ACSPhotonics}. 
In addition, it was recently reported that braiding happens in electrons with strain and phase transitions \cite{Viktor_arXiv_2021,Chen_arXiv_2021}.

\subsection{Phononic crystals}
\label{section:phononic}
Phononic crystals, also known as acoustic crystals, are also a good platform to demonstrate the topological physics because the couplings between meta-atoms can be easily controlled and the displacement field of the modes can be easily measured giving the full profile of the modes and band structure.

A nodal chain by the phononic wave was theoretically proposed using a granular metamaterial \cite{Merkel_CommPhys_2019}. For the beads consisting a face-centered cubic arrangement, one can assume that tension, shear, bending, and torsion exist on all contacts between any two nearest-neighbors grains, denotes as $K_N$, $K_S$, $G_B$, and $G_T$, respectively, in Figure~\ref{fig:PnononNodalRing}a. By applying these assumptions to the linear equations of motion for each bead, the result in Figure~\ref{fig:PnononNodalRing}b clearly shows the nodal chain in momentum space.

W. Deng et al. proposed a phononic crystal that exhibits nodal rings in momentum space \cite{Deng2019}. The effective Hamiltonian for this realization is written as
\begin{equation}
    H = d_x \sigma_1 + d_y \sigma_2
    \label{eq_Phonon_Hamiltonian}
\end{equation}
where
\begin{equation}
    d_x = -2 \operatorname{cos} k_x -2 \operatorname{cos} k_y -2 t \operatorname{cos} k_z
    \label{eq_Phonon_dx}
\end{equation}
and
\begin{equation}
    d_y = -2 \delta t \operatorname{sin} k_z
    \label{eq_Phonon_dy}
\end{equation}
are the sublattice pseudospins. For $t>0$ and $\delta t > 0$, the hopping amplitude in the $z$-direction is $-t \pm \delta t$ as illustrated in Figure~\ref{fig:PnononNodalRing}c by thin and thick vertical rods. The hopping in the $x$- and $y$-directions are $-1$ as represented in Figure~\ref{fig:PnononNodalRing}c. 
The eigenvalue of Eq.~(\ref{eq_Phonon_Hamiltonian}) is $E = \pm \sqrt{d_x ^2 + d_y ^2}$, and the degeneracies are formed when $d_x = d_y = 0$. This indicates that the pseudospins are arranged as that corresponds to nodal rings formed on the $k_z = n \pi$ plane ($n$ is an integer).

For experimental observation of the nodal rings, they prepared a layer-stacked-phononic crystal made of plastic stereolithography material. In the three-dimensional structures shown in Figure~\ref{fig:PnononNodalRing}d, there are several types of holes, the propagation paths of sound waves. Along the $z$-direction, two types of holes exist. The smaller and larger holes correspond to the thinner and thicker rods in Figure~\ref{fig:PnononNodalRing}c. 
Along the horizontal directions, the sound waves meet the same types of rectangular holes, and this corresponds to the same size of rods in Figure~\ref{fig:PnononNodalRing}c. 
With this phononic crystal design, the nodal rings could be observed in momentum space (see Figure~\ref{fig:PnononNodalRing}e and f), as predicted by Eq.~(\ref{eq_Phonon_Hamiltonian}).

\begin{figure}
    \centering
    \includegraphics{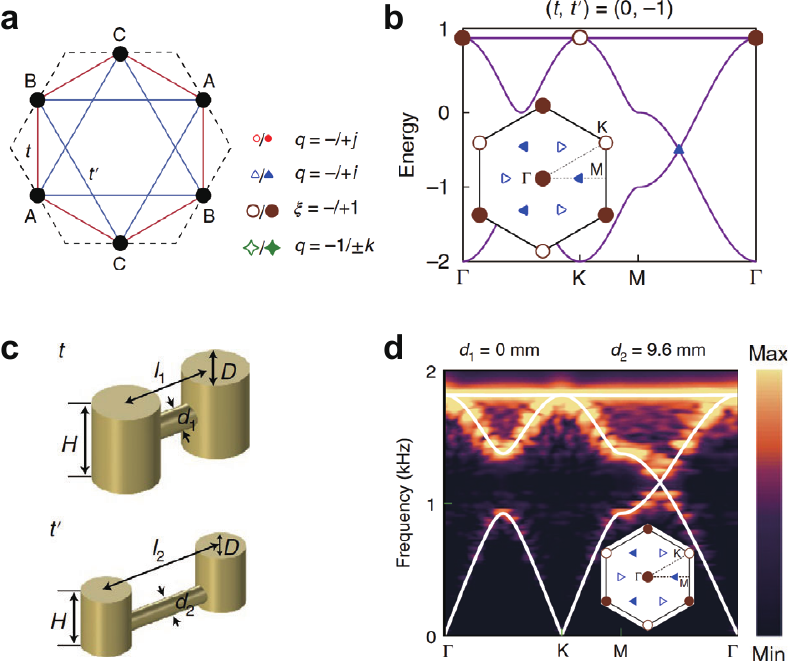}
    \caption{
        \label{fig:PhononPatchEulerClass}    
            Experimental realization of Euler class using a phononic crystal. (a) Schematics of a kagome tight-binding model with tunning parameters $t$ and $t'$. (b) Phononic band structure of (a). (c) Experimental design of (a). (d) Simulation (white plots) and experimental band structure of (c). Reproduced with permission from Ref.~\cite{Jiang2021}. Copyright 2021, Nature Portfolio (a-d).
    }
\end{figure}

Very recently, B. Jiang et al. used a two-dimensional phononic crystal to observe non-Abelian topological charges and topological phase transitions \cite{Jiang2021}. Although this work is not directly related to nodal lines in three-dimensional momentum space, this work has its significance in experimental demonstration of the non-Abelian phononic nodes (degenerate point in two-dimensional momentum space) and the patch Euler class mentioned in Section \ref{section:patchEulerClass}. 
They employed a tight-binding model of a kagome lattice as shown in Figure~\ref{fig:PhononPatchEulerClass}a. 
The hopping between lattice points A, B, and C can be adjusted by $t$ and $t'$. Band structures by tuning these two variables exhibit several types of non-Abelian charges and Euler classes, as shown in Figure~\ref{fig:PhononPatchEulerClass}b. 
Then, the tight-binding model is realized using cylindrical acoustic resonators as shown in Figure~\ref{fig:PhononPatchEulerClass}c to observe topological phase transitions and the new topological invariant ($-1$ of the Euler class) (see Figure~\ref{fig:PhononPatchEulerClass}d) .

\begin{figure*}[ht]
    \centering
    \includegraphics{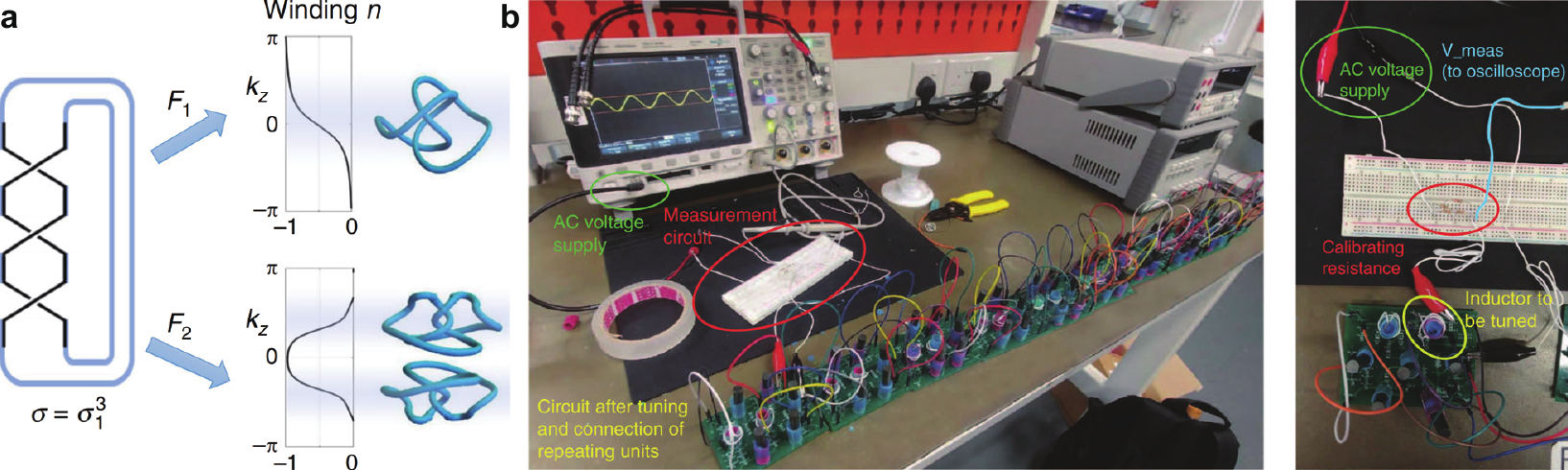}
    \caption{
        \label{fig:Topolectricity}
            (a) Mathematical models construct a nodal knot/link from a braid. A braid closure can be embedded onto the three-dimensional Brillouin zone torus differently through different choices of $F \left( \mathbf k \right)$ functions. Depending on its topological charge density distribution, it produces different numbers of nodal knots in the Brillouin zone, i.e. either a single copy ($F_1$) or two copies ($F_2$) related by mirror symmetry. (b) Experimental setup for impedance measurement of the Hopf-link circuit. Reproduced with permission from Ref.~\cite{Lee2020}. Copyright 2020, Nature Portfolio (a-b).
    }
\end{figure*}
\begin{figure*}[ht]
    \centering
    \includegraphics{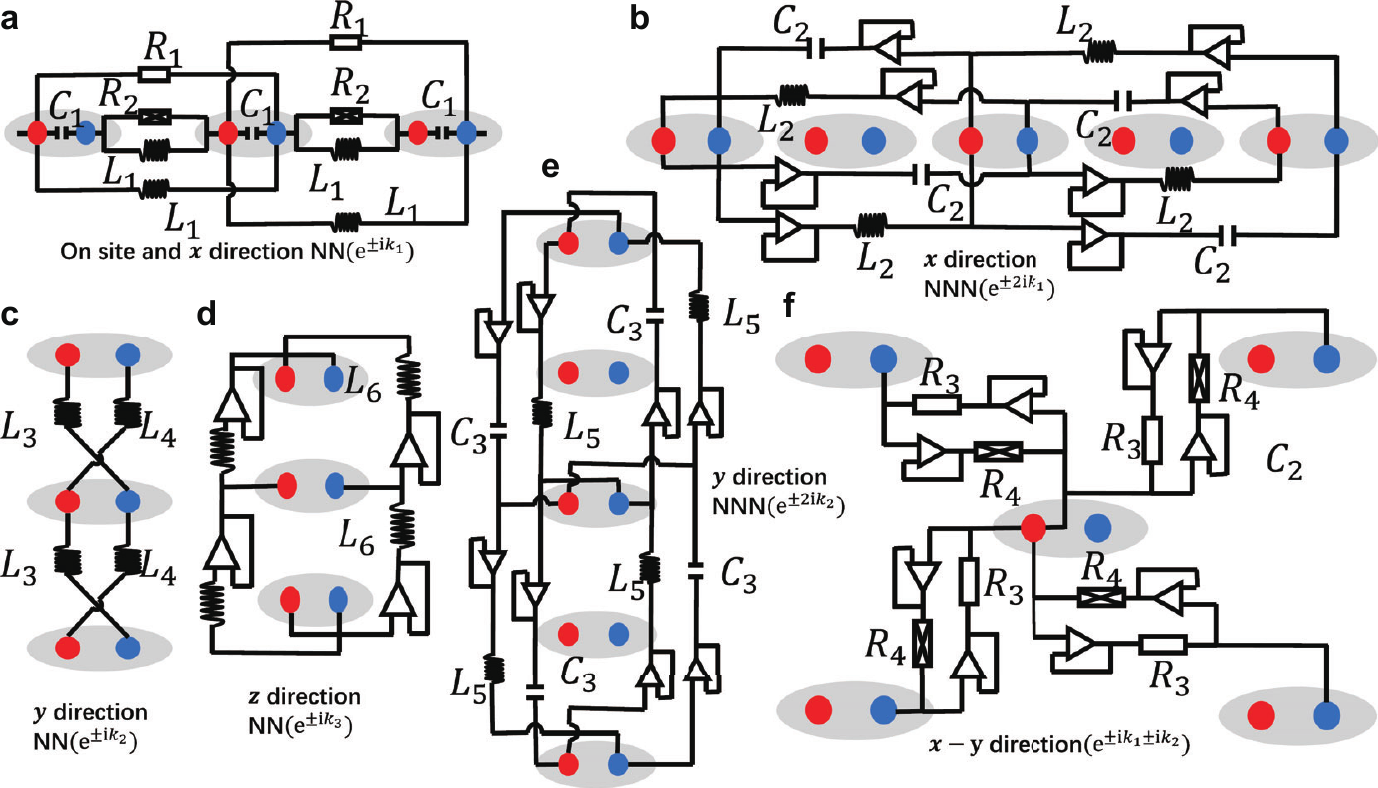}
    \caption{
        \label{fig:Non_Hermitian}    
            Illustration of different constituents of the non-Hermitian trefoil knot circuit. (a) On-site hopping and nearest-neighbor hopping along the $x$ direction, (b) next-nearest-neighbor hopping along the $x$ direction, (c) nearest-neighbor hopping along the $y$ direction, (d) nearest-neighbor hopping along the $z$ direction, (e) next-nearest-neighbor hopping along the $y$ direction, and (f) diagonal hopping within the $x-y$ plane. Reproduced with permission from Ref.~\cite{Zhang_CommPhys_2021}. Copyright 2021, Nature Portfolio (a-f).
    }
\end{figure*}

\subsection{Electrical circuits}
\label{section:electrical}
Knots, such as everyday-life ropes are intricate nodal lines. They are difficult to construct because they require finely tuned long-ranged hoppings. To realize models with those long-ranged hoppings, it is naturally suggested to use artificial structures, which allow for unprecedented control over individual couplings \cite{Lee2020}. Most importantly, electrical circuits, whose connections transcend locality and dimensionality constraints, put the implementation of couplings between distant sites of a high-dimensional system and nearest-neighbor connections on equally accessible footing. This advantage is found to be crucial to the realization of nodal knots, which contains many strong long-ranged hoppings. 

Having explained the necessity to implement nodal knots using electrical circuits, we now describe how they can be concretely implemented and detected. An electrical $RLC$ circuit is an undirected network with nodes $\alpha=1,\cdots,N$ connected by resistors, inductors, capacitors or combinations of them. Its behavior can be fully characterized by Kirchhoff's law as $I_\alpha=J_{\alpha\beta}V_\beta$, where $I_\alpha$ is the external current at junction $\alpha$ and $V_\beta$ is the voltage at junction $\beta$. Each entry $J_{\alpha\beta}$ of the Laplacian $J$ contributes $r_{ab}$ to the Laplacian, where $r_{ab}=R,i\omega L, (i \omega C)^{-1}$ for a single resistor, inductor and capacitor and whose combinations follows. The strictly reciprocal (symmetric) form of $J_{\alpha\beta}$ constrains the possible Laplacian, which prevents nodal knot circuits from being developed using mathematical models of nodal knots proposed before, since those imply broken reciprocity. Thus in order to construct nodal knot circuits, new models which preserve reciprocity need to be invented \cite{Lee2020}.

A very recent work discovered a method to overcome this obstacle, which goes beyond existing approaches that requires broken reciprocity, by pairing nodal knots with their mirror images to realize pairs of nodal knots in a fully reciprocal setting \cite{Lee2020}. The key insight is that pairs of nodal knots can preserve reciprocity while a single knot cannot, such that they can be realized in a circuit as shown in Figure~\ref{fig:Topolectricity}a. A highlight of this work is the experimental verification of surface drumhead states in a design of the nodal Hopf link circuit shown in Figure~\ref{fig:Topolectricity}b. This experimental setting is used to extract the admittance band structure through linearly independent measurements, whereas the number of measurements needed is the same as the number of inequivalent nodes $N$. Each step consists of a local excitation in this circuit and a global measurement of the voltage response from which one can extract all components of the Laplacian in reciprocal space. By diagonalizing the Laplacian $J(k)$, the admittance band structure can be found, reflecting that the Laplacian plays the same role as the Hamiltonian in an electronic material.  

There are other advantages that we have not mentioned so far in using electronic circuits to construct nodal knot as well as other nodal lines. Take the non-Hermitian nodal knots as an example, the positive, negative and non-reciprocal couplings needed for non-Hermitian nodal knots can be realized through carefully chosen combinations of $RLC$ components and operation amplifiers, which can introduce non-reciprocal feedback needed for the skin effect \cite{Zhang_CommPhys_2021} that $RLC$ components can not. By active elements in those circuits we can realize non-Hermitian models straightforwardly in electrical circuits, given the non-Hermitian trefoil circuit as an example in Figure~\ref{fig:Non_Hermitian}, which reflects the flexibility of topolectrical circuits. In those circuits, topological zero modes manifest themselves through a divergent impedance which we call topolectrical resonances. Finally, but not least, electronic circuits provide possibilities to simulate nodal lines and drumhead states beyond three dimensions, in an analogous way as simulating topological insulators in Class AI with electronic circuits in four dimensions \cite{Price_PRB_2020,Wang2020}.

\section{Conclusions and outlook} 
To summarize, we have explained the concept of nodal lines in band structures and showed that a wide range of types from a simple nodal line, a nodal link and a chain to mixed nodal lines can exist. Indeed, the extension from zero-dimensional degeneracy allows for a wider range of topology for the degeneracies in band structures of particles/waves including electrons, photons and phonons. 
Then, we have reviewed the theoretical description of the topological invariants of nodal lines with the Berry phase and the Wilczek-Zee connection based on the two seminal papers by M. V. Berry \cite{Berry_1984} and Wilczek and Zee \cite{Wilczek_PRL_1983}. This provides us with an essential toolset to describe the topology of nodal lines. Using the Wilczek-Zee connection, we explained how non-Abelian topology can be considered in a three-band system, following the recent work on non-Abelian topology by Q. Wu et al. \cite{Wu2019}. 

Finally, we have reviewed recent advances in implementing nodal lines using metamaterials and photonic crystals. These two photonic systems have successfully demonstrated those exotic behaviors and they remain very promising in finding new topological states related to nodal lines. However, they are not the only systems and they also benefit from the earlier work in other fields. To supplement, we have introduced the examples of nodal lines in electronic crystals, phononic crystals, and electrical circuits. Moreover, non-Hermitian systems, which have complex energy eigenvalues due to the exchange of energy with the environment, can be used to extend the dimension of parameters space. Indeed, new non-Hermitian systems with optical ring resonators and RLC circuits are being introduced to implement more complex topology of nodal lines such as knots and we expect many undiscovered topological nodal lines will be implemented soon with these new approaches. 

However, there are still challenges in understanding the physical consequence of topological phase of nodal lines and also implementing the proposed ideas in artificial materials. 
First, no bulk-edge/surface correspondence for the non-Abelian charges has been mathematically proven \cite{Jiang2021} although one can naturally think that there could be a relation similar to the bulk-edge/surface correspondence in Chern insulators, which is Abelian.
To answer this question,  more in-depth theoretical investigation on  multiple nodal lines systems is required. 
Second, the refractive index required to have nodal links and other topology using dielectric photonic crystals is too high, normally requiring a refractive index higher than 3.5 \cite{Park_2020_ACSPhotonics}. 
This limits the choices of materials in the microwave range, adding another challenge in fabrication. For example, L. Lu et al. used a ceramic material with a high refractive index to observe Weyl points experimentally, and they drilled the material in different directions to prepare a Weyl photonic crystal \cite{Lu_2015_Science}.
Third, the fabrication of nanostructures for dielectric photonic crystals working at optical wavelengths can be challenging. As the nodal lines are normally implemented in three-dimensional momentum space that requires a three-dimensional array of high index dielectric materials. A self-assembly method using block copolymers \cite{Jo_2021_AppliedMaterialsToday} can be used but it requires the additional step of inserting high-refractive index materials and it is hard to control the local geometrical perturbations as we want. A direct-laser writing or other polymerization methods can be used but they still require an additional step to add high-refractive index material. 
However, these challenges could be overcome in a few years considering the rapid progress in nanofabrication techniques. 

There are also unexplored areas in relation to nodal line physics in photonic systems. 
First, 3D nanoplasmonic systems can be used to implement and observe nodal lines of surface plasmon polaritons. In 2D topological photonics, there is already a theoretical study that shows unidirectional propagation and corner states in 2D metallic arrays \cite{Proctor2021}. However, due to the difficulty in fabricating 3D plasmonic structures, no nodal lines in 3D plasmonic band structures have been observed yet. Second, quantum optic systems can be used to implement the nodal line. An array of coupled quantum emitters has been studied in 2D topological photonics structures \cite{Perczel2020}. If we extend the system to 3D structures, 3D exciton polaritonic systems could be a good platform to study the topology of nodal lines. Recently, also an array of dipolar molecules in 3D optical lattices has been proposed to implement Hopf insulators \cite{Schuster2021}.

The remaining question is what applications could be enabled using nodal lines. There are very few reports or proposals regarding practical applications and it is hard to discuss a general way of applying nodal lines. 
For electronic crystals with nodal lines, high-temperature surface superconductivity using drumhead surface states \cite{Kopnin2011} has been one major motivation of study of nodal lines. Additionally, applications for surface ferromagnetism \cite{Chen2019PRL} and high harmonic generation \cite{Lee2020_HHG} have been theoretically proposed.
For photonic applications, high density of states of surface states in nodal lines systems \cite{Wang_SciRep_2021, Gao2018} are expected to enhance spontaneous emission, resonant scattering, nonlinearities and blackbody radiation. 
The bound states in the continuum related to nodal lines \cite{Yang2020} may find applications in lasing \cite{Hwang_NatComm_2021} and sensing \cite{Yanik_PNAS_2011}.
Although direct applications seem limited for the moment, a better understanding of topological phases of nodal lines would give us rich knowledge and interesting physics of artificially designed materials system as well as electronic materials. We believe this would open a new avenue to exciting applications as well as expand our human knowledge.  

\begin{acknowledgments}
The authors thank Robert-Jan Slager, Andreas Pusch and Stephan Wong for critical reading and comments.  The work is part-funded by the European Regional Development Fund through
the Welsh Government (80762- CU145 (East)). This research
is also supported by the National Natural Science Foundation of China (Grant No.11874431), the National Key R\&D Program of China (Grant No. 2018YFA0306800),
and the Guangdong Science and Technology Innovation Youth Talent Program (Grant No.2016TQ03X688).

\end{acknowledgments}




\bibliography{nodal_line}

\end{document}